\documentclass[journal,twocolumn]{IEEEtran}
\usepackage[utf8]{inputenc}
\usepackage{eqnarray,amssymb,amsmath,amsthm,mathtools,amsfonts,etoolbox}
\usepackage{mathrsfs}
\usepackage{algorithm}
\usepackage{algorithmicx}
\usepackage{algpseudocode}
\usepackage{array}
\usepackage{graphicx}
\usepackage{mdwmath}
\usepackage{mdwtab}
\usepackage{eqparbox}
\usepackage{url}
\usepackage{comment}
\usepackage{subfigure}
\usepackage{bbold}
\usepackage{color}
\usepackage{optidef}
\usepackage{makecell}
\usepackage{cite}
\usepackage{hyperref}

\algnewcommand{\Initialize}[1]{%
  \State \textbf{Initialize:}
  \Statex \hspace*{\algorithmicindent}\parbox[t]{.8\linewidth}{\raggedright #1}
}

\usepackage[english]{babel}
\newtheorem{theorem}{Theorem}
\newtheorem{lemma}{Lemma}
\theoremstyle{remark}

\newtheorem{assumption}{Assumption}
\newtheorem{corollary}{Corollary}

\newcommand*{\estimatesB}{\mathrel{\widehat=}}

\ifCLASSINFOpdf
\else
\fi

\begin{document}

\title{Distribution-Level AirComp for Wireless Federated Learning under Data Scarcity and Heterogeneity}

\author{Jun-Pyo~Hong,~\IEEEmembership{Member,~IEEE},
        Hyowoon~Seo,~\IEEEmembership{Senior Member,~IEEE},
        and Kisong~Lee,~\IEEEmembership{Senior Member,~IEEE}
\thanks{Jun-Pyo~Hong is with the School of Electronics and Electrical Engineering, Hongik University, Seoul 04066, South Korea (e-mail: jp\_hong@hongik.ac.kr).}
\thanks{Hyowoon~Seo and Kisong~Lee are with the Department of Electrical and Computer Engineering, Sungkyunkwan University, Suwon 16419, South Korea (e-mail: hyowoonseo@skku.edu; kslee851105@gmail.com).}

}

\markboth{ }%
{Shell \MakeLowercase{\textit{et al.}}: Bare Demo of IEEEtran.cls for IEEE Journals}

\maketitle

\begin{abstract}
Federated Learning (FL) has gained considerable attention as a privacy-preserving and localized approach to implementing edge artificial intelligence (AI). However, conventional FL methods face critical challenges in realistic wireless edge networks, where training data is both limited and heterogeneous, often leading to unstable training and poor generalization.
To address these challenges, we propose a Bayesian wireless FL framework that captures model uncertainty via posterior distributions and performs distribution-level aggregation, mitigating local overfitting and client drift. However, this formulation increases communication overhead and prevents the direct use of conventional Over-the-Air Computation (AirComp), which is widely used to improve communication efficiency in standard FL. To overcome this, we develop a transmission-compatible reformulation of posterior aggregation that enables distribution-level Bayesian updates to be computed over the air, 
along with a closed-form distributed transmit power control strategy derived from convergence analysis under practical wireless impairments.
Extensive simulations demonstrate that the proposed framework significantly improves test accuracy and calibration performance compared to conventional FL methods, particularly in data-scarce and heterogeneous environments.
\end{abstract}


\IEEEpeerreviewmaketitle

\section{Introduction}
\IEEEPARstart{F}{ederated} learning (FL) has emerged as a promising privacy-preserving paradigm for collaboratively training machine learning models across decentralized data sources. Instead of transmitting raw data to a central server, FL allows clients to retain their local data and only exchange model updates, thereby mitigating privacy risks \cite{TQuek2026}. A widely adopted baseline in this framework is Federated Averaging (FedAvg) \cite{McMahan2017}, where clients iteratively train local models and upload them for global averaging. Despite its simplicity and scalability, practical FL systems often suffer from statistical heterogeneity, where client data are typically non-identically distributed (non-i.i.d.) and imbalanced, reflecting personal usage patterns or localized sensing environments.

This inherent heterogeneity induces client drift, where local updates diverge from the global objective, particularly when local models are trained for multiple steps without coordination. Consequently, the aggregated model may converge slowly or even degrade in performance compared to centralized training \cite{Zhao2018_noniid}. To counteract this, various extensions to FedAvg have been proposed. For example, FedProx \cite{Smith2020_FedProx} introduces a proximal regularization term to restrict local model deviation, and SCAFFOLD \cite{Theertha2020_SCAFFOLD} employs control variates to reduce variance in gradient estimation. FedNova \cite{Poor2020_FedNova} normalizes update lengths to account for client-specific training effort, while MOON \cite{Song2021_MOON} and FedDyn \cite{Saligrama2021_FedDyn} adopt contrastive and dynamic regularization, respectively.
FedSAM \cite{FedSAM}, FedSMOO \cite{FedSMOO}, and FedGMT \cite{li2025fedgmt} introduce smoothing and sharpness-aware minimization (SAM) mechanisms, respectively, to improve the stability of local updates.

These methods effectively mitigate client drift and improve fairness across clients. However, they implicitly assume that each client holds a sufficiently large and representative dataset. In domains such as healthcare, biomedical sensing, smart homes, robotics, and cybersecurity, this assumption often breaks down. Clients may have access to only a handful of samples, and their data distributions can vary drastically. Under such data scarcity and strong non-i.i.d. conditions, local models exhibit high variance, exacerbating drift and leading to unstable or overfitted global models \cite{kamp2023_SmallData}. 

Bayesian FL has recently emerged as a principled approach to address these challenges. Unlike deterministic methods, Bayesian FL maintains a distribution over model weights rather than a deterministic point estimate, enabling explicit modeling of epistemic uncertainty across clients and during aggregation \cite{cao2023bayesianfederatedlearningsurvey}. Each client estimates a local posterior over model weights using its data, and the server aggregates these posteriors into a global one. Notably, \cite{Shao2024_BayesianFL} proposes forming the global posterior by multiplying local posteriors, yielding a distribution that minimizes the average Kullback-Leibler (KL) divergence from all local posteriors. This posterior conflation captures the diversity of learned models, regularizes training under limited data, and improves generalization in heterogeneous settings.
Similar aggregation principles have appeared in Bayesian model averaging \cite{BMA1999} and distributed Bayesian inference methods such as the Bayesian committee machine \cite{BCM2000}. However, these approaches typically perform a one-shot combination of independently trained models, whereas in FL the aggregated model is repeatedly used as the prior for subsequent rounds of local training.

Despite its theoretical advantages, Bayesian FL poses significant practical challenges. 
In particular, each client must transmit richer information, typically the mean and variance of its posterior, effectively doubling the communication payload compared to standard FL.
This overhead becomes prohibitive in wireless edge networks, where communication bandwidth is limited and channel conditions are dynamic. Moreover, existing Bayesian FL studies generally neglect the communication bottleneck, limiting their applicability in real-world systems.

To address the communication inefficiency, over-the-air computation (AirComp) has emerged as a promising technique for scalable FL in wireless networks. AirComp leverages the waveform superposition property of wireless channels to compute functions such as model averages directly in the analog domain. 
This process inherently supports linear aggregation and avoids the need for orthogonal transmissions, thereby reducing bandwidth consumption, especially in dense client scenarios \cite{Gundz2020_AirComp, Huang2020_AirComp}. Enhancements such as power control and quantization techniques have further improved AirComp performance in practical deployments \cite{JPHong2023_OTA_FL,ICC2023, Cui2022_AirComp, KHuang2020_AirCompPowOpt, KHuang2021_OneBitAirComp, kim2024}.
Recent studies have also explored digital AirComp, where coded transmissions are employed to compute aggregated functions over the air. These schemes provide improved robustness against synchronization errors and channel estimation inaccuracies through structured coding, at the cost of increased communication overhead, additional complexity, and longer transmission latency \cite{Razavikia2025, Qiao2024}.

Existing AirComp FL schemes have primarily focused on deterministic model updates (e.g., gradients or weights), and their integration with Bayesian FL remains largely unexplored. 
Some recent works \cite{gafni2023federatedlearningheterogeneousdata, NLee_2021BayesianAirComp, Gundz2023_BayesianAirComp} have employed techniques incorporating Bayesian elements for denoising or regularization in analog aggregation; however, they do not perform posterior inference and therefore do not constitute Bayesian learning in the usual sense.

In this paper, we propose a distribution-level AirComp framework that enables communication-efficient and reliable Bayesian FL in wireless edge networks characterized by non-i.i.d. and data-scarce edge device datasets. 
Unlike conventional FL, which aggregates point estimates of model weights, our framework aggregates their posterior distribution.
We consider a practical setting where each device performs variational inference and maintains a Gaussian posterior, enabling tractable distribution-level aggregation via posterior conflation.

To implement this aggregation via AirComp, each device transmits its posterior sufficient statistics in a form compatible with transmission, and the server updates the global variational distribution through appropriate normalization of the superimposed signals.
This communication–learning co-design retains uncertainty quantification while enabling efficient over-the-air aggregation.
Through theoretical analysis and empirical validation, we demonstrate that the proposed framework maintains robust learning performance under severe data heterogeneity and scarcity, while offering substantial communication gains over existing FL methods.

In summary, our key contributions are as follows:
\begin{itemize}
    \item \emph{AirComp for Communication-Constrained Bayesian Aggregation}: 
        We propose a novel AirComp framework that enables \emph{distribution-level} aggregation for Bayesian FL in wireless environments with limited communication resources.
        Unlike conventional FL, where the global model is obtained by averaging model weights, Bayesian FL requires the aggregation of additional statistics that capture the uncertainty of locally trained models.
        However, the resulting global posterior update cannot be directly implemented via AirComp due to the interdependent structure of these sufficient statistics.
        To address this challenge, each device transmits two sufficient statistics that jointly represent the local model weights and their associated uncertainty in a transmission-compatible form.
        Their superposition in the wireless channel realizes an AirComp-compatible two-phase aggregation procedure motivated by \emph{product-of-Gaussians} conflation, allowing the BS to update the global variational distribution through a simple post-processing step.
        This design enables uplink communication in constant time, regardless of the number of participating devices.
        
        \item \emph{Convergence Analysis under Wireless Impairments}:  
        To the best of our knowledge, we present the first convergence analysis for Bayesian FL with AirComp, explicitly capturing the impacts of wireless fading, additive channel noise, and update distortion.  
        In particular, the analysis reveals that communication-induced distortion in the transmitted posterior statistics directly affects the global model update and alters the convergence behavior of the learning process.  
        Theoretical guarantees are provided under channel impairments and constrained transmit power, and it is shown that, after a large number of training rounds, the distortion in aggregated updates becomes the dominant factor hindering convergence.
        
        \item \emph{Optimized Power Control}:  
        Based on our convergence analysis, we develop a power control strategy to accelerate training by mitigating the communication-induced distortion in transmitted posterior statistics.  
        Specifically, each device optimizes its transmit amplitudes to minimize its local update distortion under channel fading, local data imbalance, and a limited power budget. These distributed optimizations jointly minimize a separable upper bound on the aggregated update distortion at the BS. Since the resulting per-device power control problem admits a closed-form solution, the influence of key system parameters on convergence behavior can be explicitly characterized.
    
        \item \emph{Improved Uncertainty Calibration}:  
        In scenarios where inference performance is limited by data scarcity, uncertainty calibration becomes a key metric for evaluating the reliability of predictions.  
        The proposed framework delivers \emph{well-calibrated} predictive probabilities through probabilistic inference.  
        Empirically, our models achieve lower expected calibration error (ECE) and yield reliability diagrams that are more closely aligned with ideal confidence levels, compared to conventional FL methods with AirComp.  
        The provision of reliable confidence scores facilitates principled decision-making, making our framework particularly attractive for safety-critical applications such as home security and healthcare.
\end{itemize}
    
The rest of the paper is organized as follows. Section II provides background and preliminaries for Bayesian learning and its application in the federated setting. Section III introduces the system model and motivates the proposed framework. Section IV presents the proposed AirComp framework for Bayesian FL. Section V presents the convergence analysis and an optimized power control method to accelerate training. Section VI presents the simulation results, and Section VII concludes the paper.

\section{Background \& Preliminaries}\label{section:background}
\subsection{Variational Bayesian Method}
Contrary to standard (frequentist) machine learning, which focuses on deriving the best point estimate of model weights from data, Bayesian learning treats weights as random variables and focuses on deriving their distributions using prior information and observed data.
Specifically, Bayesian learning aims to determine the posterior distribution of the model weights $\mathbf{w}\in\mathbb{R}^d$, representing updated beliefs about these weights after taking into account the evidence provided by the data.
For a given dataset $\mathcal{D}$, the posterior distribution is computed using Bayes' rule, which updates our beliefs about the model weights by combining the likelihood of the data with the prior distribution as
\begin{align}
    p(\mathbf{w}|\mathcal{D}) &= \frac{p(\mathcal{D}|\mathbf{w})p(\mathbf{w})}{p(\mathcal{D})}.
\end{align}
The posterior distribution reflects beliefs about a range of possible model weights, making Bayesian methods more robust with limited data and better suited for estimating uncertainty—challenges not directly addressed by frequentist machine learning approaches.
However, deriving the exact posterior distribution for general deep neural network (DNN) models is computationally challenging due to the high-dimensional integration required to calculate the evidence, $p(\mathcal{D})=\int p(\mathcal{D}|\mathbf{w})p(\mathbf{w})d\mathbf{w}$. 

To circumvent this challenge, several practical techniques have been developed to estimate or approximate the posterior distribution, including Monte Carlo Markov Chain (MCMC), variational inference, and Laplace approximation. Among these, we focus on employing variational inference, which enables computationally efficient and scalable local posterior approximation. 
Variational inference derives the parameterized distribution $q_{\boldsymbol{\theta}^*}(\mathbf{w})$ that most closely approximates the true posterior distribution through the following optimization:
\begin{align}
    \boldsymbol{\theta}^*&=\underset{\boldsymbol{\theta}\in\mathbb{R}^{b}}{\arg\min} ~ \text{KL}\big[ q_{\boldsymbol{\theta}}(\mathbf{w}) \Vert p(\mathbf{w}|\mathcal{D}) \big] \nonumber \\
    &= \underset{\boldsymbol{\theta}\in\mathbb{R}^{b}}{\arg\min} ~ \text{KL}\big[ q_{\boldsymbol{\theta}}(\mathbf{w}) \Vert p(\mathbf{w}) \big] -\mathbb{E}_{q_{\boldsymbol{\theta}}(\mathbf{w})}\big[ \log p(\mathcal{D}|\mathbf{w}) \big]\label{eq:TwoTermVILoss}\\
    &= \underset{\boldsymbol{\theta}\in\mathbb{R}^{b}}{\arg\min} ~ L_{\text{VI}}(\boldsymbol{\theta}), \label{eq:VI}
\end{align}
where $\text{KL}[f(\mathbf{w})\Vert g(\mathbf{w})]=\int_{\mathbf{w}} f(\mathbf{w})\log \frac{f(\mathbf{x})}{g(\mathbf{w})}$ denotes Kullback-Leibler (KL) divergence of the probability distribution $f(\mathbf{w})$ from $g(\mathbf{w})$, the log likelihood $\log p(\mathcal{D} | \mathbf{w})$ becomes $\sum_{(x,y)\in\mathcal{D}}\log p(y|x, \mathbf{w})$ for supervised learning, and $L_{\text{VI}}(\boldsymbol{\theta})=\text{KL}\big[ q_{\boldsymbol{\theta}}(\mathbf{w}) \Vert p(\mathbf{w}) \big] -\mathbb{E}_{q_{\boldsymbol{\theta}}(\mathbf{w})}\big[ \log p(\mathcal{D}|\mathbf{w}) \big]$ is referred to as the variational free energy or the negative evidence lower bound (ELBO).
To facilitate computing the expectation with respect to $q_{\boldsymbol{\theta}}(\mathbf{w})$, the negative ELBO is approximated by
\begin{align}\label{eq:approx_L}
    L_{\text{VI}}(\boldsymbol{\theta}) \approx \frac{1}{M}\sum_{m}^{M} \log q_{\boldsymbol{\theta}}(\mathbf{w}_{(m)}) - \log p(\mathbf{w}_{(m)}) - \log p(\mathcal{D}|\mathbf{w}_{(m)}),
\end{align}
where $\mathbf{w}_{(m)}$ denotes the $m$-th Monte Carlo (MC) sample drawn from the distribution $q_{\boldsymbol{\theta}}(\mathbf{w})$.
Then, stochastic gradient descent (SGD) can be applied directly to \eqref{eq:approx_L} to solve the optimization problem \eqref{eq:VI} \cite{VI_2011, VI_2015,Seo2024}.

Specifically, for a Gaussian variational distribution under a mean-field approximation, the variational parameters consist of a mean vector and a diagonal covariance matrix, $\boldsymbol{\theta}=(\boldsymbol{\mu}, \mathbf{\Sigma})$.
To compute the approximate loss in \eqref{eq:approx_L}, a MC sample $\mathbf{w}_{(m)}$ is drawn via the reparameterization $\mathbf{w}_{(m)}=\boldsymbol{\mu} + \mathbf{\Sigma}^{1/2}\boldsymbol{\epsilon}_{(m)}$, where $\boldsymbol{\epsilon}_{(m)}\sim\mathcal{N}(0, \mathbf{I}_d)$.
\begin{figure*}[t]
    \centering
    \includegraphics[width=0.8\textwidth]{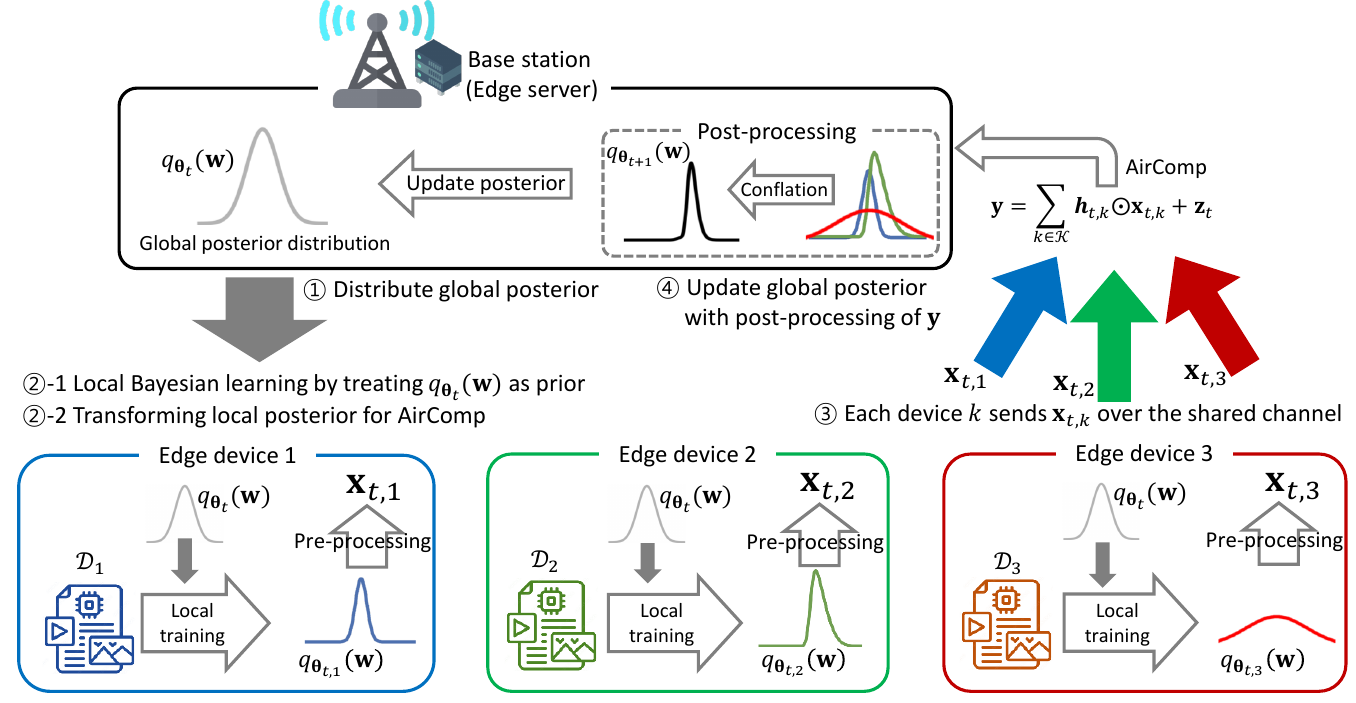}
    \caption{Wireless Bayesian FL with AirComp}\label{fig:SystemModel}
\end{figure*}
\subsection{Bayesian FL}\label{subsection:BFL}
For a given set of edge devices $\mathcal{K}=\{1, 2,...,K\}$ and their distributed local datasets $\{\mathcal{D}_1, \mathcal{D}_2, ..., \mathcal{D}_K\}$, the goal of Bayesian FL is to derive the posterior distribution of the model weights $\mathbf{w}\in\mathbb{R}^d$ toward the global dataset $\mathcal{D}=\bigcup_{k\in\mathcal{K}}\mathcal{D}_k$, denoted as $p(\mathbf{w}|\mathcal{D})$.
However, it is infeasible to directly evaluate $p(\mathbf{w}|\mathcal{D})$ with the distributed local datasets.
To address this limitation in the FL setup, the global posterior distribution can be estimated by collecting locally computed posterior distributions, $\{p(\mathbf{w}|\mathcal{D}_k)\}_{k\in\mathcal{K}}$, from devices and conflating them into a single representative distribution \cite{Shao2024_BayesianFL}. 

Suppose variational inference is employed by each device to approximate its local posterior distribution.
At the beginning of training round $t$, the server disseminates the global variational posterior distribution $q_{\boldsymbol{\theta}_t}(\mathbf{w})$ to all devices. 
Each device $k\in\mathcal{K}$ then performs local training on its dataset $\mathcal{D}_k$, treating the global posterior distribution as a prior in \eqref{eq:VI}.
In other words, the variational parameters $\boldsymbol{\theta}_{t,k}$ locally trained at device $k$ is obtained by solving the following optimization problem: 
\begin{align}
    \boldsymbol{\theta}_{t,k} &= \underset{\boldsymbol{\theta}\in\mathbb{R}^b}{\arg\min} ~ \lambda\text{KL}\!\left[ q_{\boldsymbol{\theta}}(\mathbf{w})\Vert q_{\boldsymbol{\theta}_{t}}(\mathbf{w}) \right] - \mathbb{E}_{q_{\boldsymbol{\theta}}(\mathbf{w})} \!\left[ \log p(\mathcal{D}_k|\mathbf{w}) \right]\nonumber\\
    &=\underset{\boldsymbol{\theta}\in\mathbb{R}^b}{\arg\min} ~ L_k(\boldsymbol{\theta}), \label{eq:LocalTrainProb}
\end{align}
where $L_k(\boldsymbol{\theta})=\lambda\text{KL}\!\left[ q_{\boldsymbol{\theta}}(\mathbf{w})\Vert q_{\boldsymbol{\theta}_{t}}(\mathbf{w}) \right] - \mathbb{E}_{q_{\boldsymbol{\theta}}(\mathbf{w})} \!\left[ \log p(\mathcal{D}_k|\mathbf{w}) \right]$, and $\lambda$ denotes a positive constant that regulates the degree of deviation from the global posterior distribution during local training.
After completing the local training process defined by \eqref{eq:LocalTrainProb}, each device $k$ reports its locally updated variational parameters $\boldsymbol{\theta}_{t,k}$ to the server.
The server then aggregates these updates and refines the global posterior distribution by conflating the local variational posterior distributions $\{q_{\boldsymbol{\theta}_{t,k}}(\mathbf{w})\}_{k\in\mathcal{K}}$.
In large classes of distributions, including the Gaussian distribution, the conflated distribution that minimizes Shannon information loss is formed by the normalized product of the local distributions \cite{Hill2011_Conflation}.
Assuming a Gaussian variational distribution, the global posterior distribution is updated as
\begin{align}
    p(\mathbf{w}|\mathcal{D}) &\estimatesB q_{\boldsymbol{\theta}_{t+1}}(\mathbf{w}) \nonumber\\
    &= C\prod_{k \in \mathcal{K}} q_{\boldsymbol{\theta}_{t,k}}(\mathbf{w})^{\pi_k},\label{eq:GlobalEst}
\end{align} 
where $\pi_k = |\mathcal{D}_k|/|\mathcal{D}|$, and $C$ denotes a normalization constant.
Once the global posterior distribution is refined, training proceeds to round $t+1$.
Consequently, the global posterior distribution is improved through the iterative aggregation and dissemination over multiple training rounds.

Although Bayesian approaches in \cite{Ozer2022, zhang2023BFL, Shao2024_BayesianFL} have been introduced to mitigate training performance degradation under non-i.i.d. data by incorporating probabilistic modeling, the impact of limited communication capacity on training performance and potential strategies to address this challenge remain unexplored.
Updating the global posterior distribution at the server relies on the received local variational parameters $\{\boldsymbol{\theta}_{t,k}\}_{k\in\mathcal{K}}$, making the design of efficient transmission schemes for parameter aggregation essential in practical wireless networks. Such efficiency not only reduces the required communication overhead but also accelerates training convergence, especially under constrained communication resources.

\section{System Model}
As illustrated in figure \ref{fig:SystemModel}, we consider a single cell wireless edge network with $K$ devices distributed in a circular region around a BS, serving as an edge server. We focus on challenging scenarios where the number of participating devices $K$ is large, but each device $k\in\mathcal{K}$ has a small, unbalanced, and non-i.i.d. local dataset $\mathcal{D}_k$.

Our framework is centered on designing a transmission method tailored to learning a global variational distribution that effectively represents all distributed local datasets by optimizing the global cost function defined as
\begin{align}\label{eq:GlobL}
    L(\boldsymbol{\theta}) &= \sum_{k\in\mathcal{K}} \pi_k L_k(\boldsymbol{\theta}),
\end{align}
where $L_k$ is a local cost function at device $k \in \mathcal{K}$.

The proposed Bayesian FL framework follows the local training and aggregation processes outlined in Subsection \ref{subsection:BFL}, while accounting for additional challenges arising from unstable wireless channels during the exchange of variational posterior parameters between the BS and the devices.
Communication between each device and the BS occurs over random and heterogeneous wireless channels influenced by small-scale fading and path loss. 

The variational parameters of the global posterior distribution, $\boldsymbol{\theta}_t$, are disseminated from the BS to all devices via reliable digital multicast communication at each training round $t$. Accordingly, the number of downlink transmissions required to deliver these parameters is determined by the worst-channel device. For communication-efficient uplink aggregation of local updates, we adopt AirComp with analog modulation, as described in the following subsection.

\subsection{AirComp under OFDM Fading Channels}
To mitigate the excessive resource demands caused by orthogonal channel allocation in large-scale networks, we utilize AirComp approach for local update aggregation. 
This approach allows devices to concurrently transmit analog-modulated data representing their local posterior distribution updates over a shared channel. By exploiting the superposition property of the wireless medium, the server achieves efficient aggregation of local updates from all devices in constant time, independent of the number of participating devices.

Using the locally optimized variational parameters from \eqref{eq:LocalTrainProb}, each device $k \in \mathcal{K}$ constructs $N$ OFDM symbols to transmit its local update to the BS. 
Each OFDM symbol $\mathbf {x}_{t,k}^{(n)}$ can convey up to $F$ analog values over $F$ orthogonal sub-channels, where $n\in\{1, 2, ..., N\}$.
Each OFDM symbol should satisfy the symbol power constraint 
\begin{align} \label{eq:PowerConst}
    \left\Vert\mathbf {x}_{t,k}^{(n)}\right\Vert^2\le P.
\end{align}
With the power constraint \eqref{eq:PowerConst}, the transmit power of each OFDM symbol cannot exceed $P$.
During the $n$-th transmission interval, the received signal of BS over $F$ sub-channels are represented by a vector
\begin{align}\label{eq:ReceivedSignal}
    \mathbf{y}_t^{(n)} = \sum_{k\in \mathcal{K}}\mathbf{h}_{t,k}\odot\mathbf{x}_{t,k}^{(n)} + \mathbf{z}_t^{(n)},
\end{align}
where $\mathbf{z}^{(n)}\sim\mathcal{CN}(0, \sigma_z^2\mathbf{I}_F)$ denotes an effective noise which accounts for additive noise and out-of-cell interference, $\mathbf{h}_{t,k}\in\mathbb{C}^{F}$ denotes the channel gain vector of device $k$ during the training round $t$, and the operator $\odot$ denotes the Hadamard product.
We consider a Rayleigh block fading channel, where each channel gain remains constant for at least $N$ OFDM symbols and varies independently across training rounds.
Specifically, the entries of the channel gain vector are assumed to be i.i.d. complex Gaussian random variables, $\mathbf{h}_{t,k}\sim\mathcal{CN}\left(0, r_k^{-\alpha_\text{PL}}\mathbf{I}_F\right)$ for $k\in \mathcal{K}$.
The channel variance $r_k^{-\alpha_\text{PL}}$ accounts for the propagation path-loss from device $k$ to BS, where $r_k$ and $\alpha_\text{PL}$ denote the link distance and path-loss exponent, respectively.
Since the dissemination of the global Bayesian model and the collection of local updates occur over wireless channels in a time-division duplex (TDD) manner, each device is assumed to know its local channel state information (CSI).
On the other hand, the BS is assumed not to know the CSI of each individual device due to the inordinate overhead for the global CSI acquisition in large scale networks.

Note that since local update aggregation in Bayesian FL \eqref{eq:GlobalEst} is not simply an averaging operation as it is in conventional FL, typical AirComp approaches, which aim to directly average local parameters or updates, cannot be applied to update the global posterior distribution.
Furthermore, because channel fading and additive noise introduce estimation errors at the BS, the transmit signal $\mathbf{x}_{t,k}$ should be designed to mitigate the effects of these channel impairments while leveraging the benefits of AirComp to accelerate training convergence.

\section{Bayesian FL with Over-the-Air Posterior Aggregation}
\subsection{Gaussian Variational Inference for Local Training}
For tractable analysis, we restrict the variational distribution to a Gaussian under the mean-field approximation \cite{VI_2015, Shao2024_BayesianFL}.
Hence, describing the variational posterior distribution explicitly requires twice as many parameters as the model size $d$.
We define the variational parameters $\boldsymbol{\theta}\in\mathbb{R}^{2d}$ as 
\begin{align}
    \boldsymbol{\theta} = f(\boldsymbol{\mu}, \mathbf{\Sigma}),
\end{align}
where $f:\mathbb{R}^{2d}\rightarrow \mathbb{R}^{2d}$ denotes an invertible deterministic function, and $\boldsymbol{\mu}\in\mathbb{R}^d$ and $\mathbf{\Sigma}\in\mathbb{R}^{d\times d}$ are respectively the mean vector and diagonal covariance matrix of the model weights $\mathbf{w}\in\mathbb{R}^d$.

Based on Gaussian local and global variational distributions, the computational complexity involved in calculating the KL divergence in \eqref{eq:VI} can be reduced by applying lemma \ref{lemma:KL_gauss}.
\begin{lemma}\label{lemma:KL_gauss}
    The KL divergence between two multivariate Gaussian distributions, from $q_1(\mathbf{w})=\mathcal{N}(\mathbf{w}| \boldsymbol{\mu}_1, \mathbf{\Sigma}_1)$ to $q_2(\mathbf{w})=\mathcal{N}(\mathbf{w}| \boldsymbol{\mu}_2, \mathbf{\Sigma}_2)$, has a closed-form expression given by
    \begin{align}
        \text{KL}[q_1\Vert q_2] &= \frac{1}{2} \bigg( \log\frac{|\mathbf{\Sigma}_2|}{|\mathbf{\Sigma}_1|} - d + \text{tr}\!\left[\mathbf{\Sigma}_2^{-1}\mathbf{\Sigma}_1\right] \nonumber\\
        &\qquad~~~ + (\boldsymbol{\mu}_2-\boldsymbol{\mu}_1)^\intercal\mathbf{\Sigma}_2^{-1} (\boldsymbol{\mu}_2-\boldsymbol{\mu}_1)\bigg). 
    \end{align}  
    \begin{proof}
    \begin{align}
        &\text{KL}[q_1\Vert q_2] \nonumber\\
        &= \mathbb{E}_{q_1} \!\bigg[ \frac{1}{2}\log\frac{|\mathbf{\Sigma}_2|}{|\mathbf{\Sigma}_1|} +\frac{1}{2}(\mathbf{w}-\boldsymbol{\mu}_2)^\intercal\mathbf{\Sigma}_2^{-1} (\mathbf{w}-\boldsymbol{\mu}_2)\nonumber\\
        &\qquad~~~ 
        - \frac{1}{2}(\mathbf{w}-\boldsymbol{\mu}_1)^\intercal\mathbf{\Sigma}_1^{-1} (\mathbf{w}-\boldsymbol{\mu}_1) \bigg] \nonumber\\
        &= \frac{1}{2}\log\frac{|\mathbf{\Sigma}_2|}{|\mathbf{\Sigma}_1|} +\frac{1}{2}\mathbb{E}_{q_1} \!\bigg[(\mathbf{w}-\boldsymbol{\mu}_2)^\intercal\mathbf{\Sigma}_2^{-1} (\mathbf{w}-\boldsymbol{\mu}_2) \bigg] \nonumber\\
        &~~~ - \frac{1}{2}\text{tr}\!\left[\mathbb{E}_{q_1}\! \left[(\mathbf{w}-\boldsymbol{\mu}_1)(\mathbf{w}-\boldsymbol{\mu}_1)^\intercal\right]\mathbf{\Sigma}_1^{-1}\right]\nonumber\\
        &= \frac{1}{2} \bigg(\! \log\frac{|\mathbf{\Sigma}_2|}{|\mathbf{\Sigma}_1|} \!+\! \text{tr}\!\left[\mathbf{\Sigma}_2^{-1}\mathbf{\Sigma}_1\right] \!+\! (\boldsymbol{\mu}_2\!-\!\boldsymbol{\mu}_1)^\intercal\mathbf{\Sigma}_2^{-1} (\boldsymbol{\mu}_2\!-\!\boldsymbol{\mu}_1) \!-\! d \bigg).
    \end{align}        
    \end{proof}
\end{lemma}
We define the closed-form expression of the first term in \eqref{eq:TwoTermVILoss} as a regularization loss 
\begin{align}\label{eq:KL_div}
    &L_{\text{reg}, t}(\boldsymbol{\theta}) = \text{KL}\big[ q_{\boldsymbol{\theta}}(\mathbf{w}) \Vert q_{\boldsymbol{\theta}_{t}}(\mathbf{w}) \big]\nonumber\\
    &=\frac{1}{2} \bigg(\! \log\frac{|\mathbf{\Sigma}_{t}|}{|\mathbf{\Sigma}|} \!+\! 
 \text{tr}\!\left[\mathbf{\Sigma}_{t}^{-1}\mathbf{\Sigma}\right]  +\! (\boldsymbol{\mu}_{t}\!-\!\boldsymbol{\mu})^\intercal\mathbf{\Sigma}_{t}^{-1} (\boldsymbol{\mu}_{t}\!-\!\boldsymbol{\mu}) - d\bigg).
\end{align}
To address the computational challenges associated with evaluating the second expectation term in \eqref{eq:TwoTermVILoss}, we introduce a task loss function that approximates this term using Monte Carlo (MC) sampling
\begin{align}\label{eq:task_loss}
    L_{\text{task}, k}(\boldsymbol{\theta}) = \frac{1}{M}\sum_{m=1}^{M}  - \log p\big(\mathcal{D}_k|\mathbf{w}_{(m)}\big),
\end{align}
where $\mathbf{w}_{(m)}$ denotes the $m$-th MC sample drawn from the variational distribution $q_{\boldsymbol{\theta}}(\mathbf{w})$, and $M$ denotes the number of MC samplings.
Accordingly, in training round $t$, each device $k\in\mathcal{K}$ derives the optimized variational posterior distribution by employing the SGD algorithm on the local cost function
\begin{align}\label{eq:localLoss}
    L_{k}(\boldsymbol{\theta}) = L_{\text{task}, k}(\boldsymbol{\theta}) + \lambda_k L_{\text{reg}, t}(\boldsymbol{\theta}),
\end{align} 
where $\lambda_k=\lambda \frac{|\mathcal{D}_k|}{|\mathcal{D}|d}=\lambda \frac{\pi_k}{d}$ for a scaling constant $\lambda$. 
The coefficient $\lambda_k$ is designed to compensate for the scaling mismatch between the task loss $L_{\text{task}, k}(\boldsymbol{\theta})$ and the regularization loss $L_{\text{reg}, t}(\boldsymbol{\theta})$ in the local training objective, namely the negative ELBO. This is because, as shown in \eqref{eq:KL_div} and \eqref{eq:task_loss}, the KL divergence term scales with the number of parameters $d$, whereas the log-likelihood term scales with the local dataset size $|\mathcal{D}_k|$.

Eventually, local training aims to derive the local variational parameters $\boldsymbol{\theta}_{t,k}$, which minimize the local cost function \eqref{eq:localLoss} for a given dataset $\mathcal{D}_k$.
The local variational distribution is given by
\begin{align}
    q_{\boldsymbol{\theta}_{t,k}}(\mathbf{w}) = \mathcal{N}\!\left(\mathbf{w}| \boldsymbol{\mu}_{t,k}, \mathbf{\Sigma}_{t,k}  \right),    
\end{align}
where $\boldsymbol{\mu}_{t,k} \in \mathbb{R}^{d}$ and $\mathbf{\Sigma}_{t,k}\in \mathbb{R}^{d\times d}$ are respectively the mean vector and the diagonal covariance matrix  obtained from local training,
and $\boldsymbol{\theta}_{t,k}=f\!\left(\boldsymbol{\mu}_{t,k}, \mathbf{\Sigma}_{t,k}\right)$.

\subsection{AirComp for Posterior Aggregation} 
Given that the product of Gaussian distributions yields a Gaussian-shaped distribution, the conflated global variational distribution also takes the form of a Gaussian distribution. 
From \eqref{eq:GlobalEst} and $\{(\boldsymbol{\mu}_{t,k}, \mathbf{\Sigma}_{t,k})\}_{k\in\mathcal{K}}$, the global variational distribution is given by  
\begin{align}
    q_{\boldsymbol{\theta}_{t+1}}(\mathbf{w}) &= C\prod_{k\in\mathcal{K}} \mathcal{N}(\mathbf{w}|\boldsymbol{\mu}_{t,k}, \mathbf{\Sigma}_{t,k})^{\pi_k} \nonumber\\
    &= \mathcal{N}\!\left( \mathbf{w} |  \boldsymbol{\mu}_{t+1}, \mathbf{\Sigma}_{t+1} \right),\nonumber
\end{align}
where 
$\boldsymbol{\theta}_{t+1} = f\!\left(\boldsymbol{\mu}_{t+1}, \mathbf{\Sigma}_{t+1}\right)$, and mean and diagonal covariance matrix are characterized with
\begin{align}
    \mathbf{\Sigma}_{t+1}^{-1} &= \sum_{k\in\mathcal{K}} \pi_k \mathbf{\Sigma}_{t,k}^{-1} \label{eq:Sigma}\\
    \boldsymbol{\mu}_{t+1} &= \mathbf{\Sigma}_{t+1}  \sum_{k\in\mathcal{K}} \pi_k \boldsymbol{\mu}_{t,k}\mathbf{\Sigma}_{t,k}^{-1}.\label{eq:mu_sigma}
\end{align}
From the relationship between the local and global variational distributions in \eqref{eq:Sigma} and \eqref{eq:mu_sigma}, the aggregation of the local variational posteriors inherently involves an additive structure across devices.
This property enables the use of AirComp, with appropriate pre- and post-processing, to reduce the communication resources required for aggregation in Bayesian FL.
Furthermore, since the update of the global mean $\boldsymbol{\mu}_{t+1}$ depends on the updated covariance matrix $\mathbf{\Sigma}_{t+1}$, the covariance in \eqref{eq:Sigma} must be obtained prior to performing the mean update in \eqref{eq:mu_sigma}.

Motivated by the coupled precision-weighted structure of Gaussian conflation in \eqref{eq:Sigma} and \eqref{eq:mu_sigma}, we propose an AirComp compatible two-phase aggregation procedure. The first phase aggregates the local precision updates, whereas the second phase uses the resulting global covariance as a common approximation when updating the transformed mean parameters.
Based on \eqref{eq:Sigma} and \eqref{eq:mu_sigma}, we define variational parameters for device $k$ at training round $t$ as 
\begin{align}
    \boldsymbol{\theta}_{t,k}=\left[\begin{array}{l}\boldsymbol{\rho}_{t,k}\\ \boldsymbol{\nu}_{t,k}\end{array}\right], \nonumber
\end{align}
where 
\begin{align}
    \boldsymbol{\rho}_{t,k} = \text{diag}\!\left( \mathbf{\Sigma}_{t,k}^{-1} \right),\label{eq:rho_def}\\
    \boldsymbol{\nu}_{t,k} = \mathbf{\Sigma}_{t+1}\mathbf{\Sigma}_{t,k}^{-1}\boldsymbol{\mu}_{t,k}. \label{eq:weighting_agg}
\end{align}

Substituting \eqref{eq:rho_def} and \eqref{eq:weighting_agg} into \eqref{eq:mu_sigma} and \eqref{eq:Sigma}, the aggregation can be rewritten in a linear form suitable for AirComp as follows:
\begin{align}
    \text{diag}\big(\mathbf{\Sigma}_{t+1}^{-1}\big) &= \sum_{k\in\mathcal{K}} \pi_k\boldsymbol{\rho}_{t,k}, \label{eq:precisionUpdate}\\
    \boldsymbol{\mu}_{t+1} &= \sum_{k\in\mathcal{K}} \pi_k \boldsymbol{\nu}_{t,k}. \label{eq:mu_update}
\end{align}
Accordingly, \eqref{eq:precisionUpdate} and \eqref{eq:mu_update} preserve the precision-weighted aggregation structure motivated by Gaussian conflation, but constitute a two-phase approximation rather than an exact joint implementation of \eqref{eq:Sigma} and \eqref{eq:mu_sigma}.
This formulation provides an intuitive interpretation of the aggregation process. In particular, \eqref{eq:rho_def} reflects the confidence of each local estimate through the inverse covariance, where a larger value indicates higher reliability. 
Accordingly, devices with higher confidence contribute more to the global covariance update, as reflected in \eqref{eq:precisionUpdate}.
Building on this, \eqref{eq:weighting_agg} incorporates this confidence into the local mean, effectively weighting each estimate according to its reliability.
Consequently, the aggregation process inherently assigns greater importance to more reliable local estimates, leading to a global model that reflects the collective confidence across devices while reducing the influence of more uncertain ones.

Specifically, at the beginning of the first phase, a new variational parameter vector $\boldsymbol{\rho}_{t,k}$ is defined for each device $k\in\mathcal{K}$, initialized as 
\begin{align}\label{eq:rho_init}
    \boldsymbol{\rho}_{t,k} \leftarrow \text{diag}\big( \mathbf{\Sigma}_t^{-1}\big).
\end{align}
By applying the reparameterization trick with \eqref{eq:rho_init}, each device $k$ generates MC samples from $\mathcal{N}(\boldsymbol{\mu}_t, \mathbf{\Sigma}_{t,k})$ according to 
\begin{align}
    \mathbf{w}_{(m)} = \boldsymbol{\rho}_{t,k}^{-\frac{1}{2}}\odot\boldsymbol{\epsilon}_{(m)} + \boldsymbol{\mu}_t,
\end{align}
where $\boldsymbol{\epsilon}_{(m)}\in\mathbb{R}^{d}$ is the $m$-th MC sample from $\mathcal{N}(0, \mathbf{I}_d)$, and $\boldsymbol{\rho}_{t,k}^{-\frac{1}{2}}$ is the element-wise inverse square root of $\boldsymbol{\rho}_{t,k}$. 
The local cost function $L_k(\boldsymbol{\mu}_t, \boldsymbol{\rho}_{t,k})$ is computed using $M$ independent MC samples, and a gradient descent step for the variational parameter vector $\boldsymbol{\rho}_{t,k}$ is given by
\begin{align}\label{eq:update_phase1}
    \boldsymbol{\rho}_{t,k} \leftarrow \boldsymbol{\rho}_{t,k} - \eta \nabla_\rho L_k\!\left(\boldsymbol{\mu}_t, \boldsymbol{\rho}_{t,k}  \right).
\end{align}
The local update of $\boldsymbol{\rho}_{t,k}$ is performed over $E$ gradient descent steps.

After the local training, the local update vector of $\boldsymbol{\rho}_{t,k}$ is represented as
\begin{align}\label{eq:rho_normal}
    \mathbf{\Delta}_{\rho_{t,k}} = \boldsymbol{\rho}_{t,k}-\boldsymbol{\rho}_{t}.
\end{align}
The update vector $\mathbf{\Delta}_{\rho_{t,k}}$ is partitioned into $N_1$ groups based on a predefined grouping rule.
Then, the update values in each group is transmitted over $F$ sub-channels of an OFDM symbol, represented by the vector $\mathbf{\Delta}_{\rho_{t,k}}^{(n_1)}\in\mathbb{R}^F$ for OFDM symbol $n_1\in\{1, 2, ..., N_1\}$. 
To accommodate cases where the number of values in a group is less than $F$, zeros are padded, ensuring that $\mathbf{\Delta}_{\rho_{t,k}}^{(n_1)}$ forms a $F$-dimensional vector.
It is important to note that each device transmits only the update vector rather than the variational parameters themselves. Since these updates correspond to incremental changes produced by gradient-based local training, their magnitudes are typically much smaller than those of the variational parameters, which significantly reduces the dynamic range of the transmitted signals.

In AirComp transmissions, it is crucial for each device to carefully control its transmit power to ensure that the received signal strengths at the BS are aligned, even under varying channel fading conditions across devices.
To achieve this, the BS first calculates the average update power $\bar{\delta}_{\rho_t}$ by collecting the local update powers $\frac{1}{d}\Vert\mathbf{\Delta}{\rho_{t,k}} \Vert^2$ from all devices and then computing their weighted average as follows
\begin{align}\label{eq:rho_avg}
    \bar{\delta}_{\rho_t} = \sum_{k\in\mathcal{K}} \frac{\pi_k}{d}\Vert\mathbf{\Delta}_{\rho_{t,k}} \Vert^2.
\end{align}
The BS then sends $\bar{\delta}_{\rho_t}$ back to all devices.
We assume these magnitudes are delivered without errors and require only negligible communication resources, since each is a single scalar value.
Using this global statistic $\bar{\delta}_{\rho_t}$, each device scales its local update so that all devices transmit their updates with a consistent reference scale.
Each device $k$ then computes its power allocation vectors $\mathbf{p}_{t,k}^{(n_1)}\in\mathbb{R}^{F}$ for symbols $n_1\in\{1, 2, ..., N_1\}$ according to 
\begin{align}\label{eq:PowerAlloc_1}
    \mathbf{p}_{t,k}^{(n_1)} &= \pi_k^2\gamma\mathbf{g}_{t,k} \odot \frac{\mathbf{v}_{\rho_{t,k}}^{(n_1)}\odot \mathbf{v}_{\rho_{t,k}}^{(n_1)}}{\bar{\delta}_{\rho_{t}}}\nonumber\\
                       &= \mathbf{u}_{t,k}\odot \left( \mathbf{v}_{\rho_{t,k}}^{(n_1)}\odot \mathbf{v}_{\rho_{t,k}}^{(n_1)}\right),
\end{align}
where $\gamma$ denotes a positive coefficient for power scaling, $\mathbf{g}_{t,k}\in\mathbb{R}^{F}$ denotes the channel power inversion whose entry $f\in \{1, 2, ..., F\}$ is defined as $g_{t,k,f}=\big|h_{t,k,f}\big|^{-2}$, and $\mathbf{v}_{\rho_{t,k}}^{(n_1)}\in\mathbb{R}^{F}$ denotes a power control vector to be optimized.
The vector $\mathbf{u}_{t,k} = \frac{\pi_k^2\gamma}{\bar{\delta}_{\rho_{t,k}}}  \mathbf{g}_{t,k}$ helps to preserve a certain level of received SNR under channel fading, while  $\mathbf{v}_{\rho_{t,k}}^{(n_1)}$ constrains the magnitudes of the entries in $\mathbf{\Delta}_{\rho_{t,k}}^{(n_1)}$ to satisfy the power constraint \eqref{eq:PowerConst}, $\big\Vert \mathbf{p}_{t,k}^{(n_1)} \big\Vert_1\le P$.
It is important to note that the channel inversion factor $\mathbf{g}_{t,k}$ in \eqref{eq:PowerAlloc_1} does not imply that full channel inversion is enforced across all subchannels. Instead, the actual transmit magnitude is jointly controlled by the power control vector $\mathbf{v}_{\rho_{t,k}}^{(n_1)}$ under the transmit power constraint.
For example, when a subchannel is severely faded with a large $g_{t,k,f}$, the power constraint can still be satisfied by reducing the corresponding entry of $v_{\rho_{t,k},f}^{(n_1)}$, which suppresses the transmitted update magnitude on that subchannel at the expense of introducing distortion to the local update.
Eventually, each device $k\in\mathcal{K}$ reports the power-constrained version of local update vector $\text{sgn}(\mathbf{\Delta}^{(n_1)}_{\rho_{t,k}})\odot\mathbf{v}_{\rho_{t,k}}^{(n_1)}$ instead of $\mathbf{\Delta}^{(n_1)}_{\rho_{t,k}}$, where $\text{sgn}(\cdot)$ denotes the element-wise signum function.
In the training round $t$, the device $k$ constitutes $n_1$-th symbol as
\begin{align}\label{eq:Gen_x}
    \mathbf{x}_{t,k}^{(n_1)} &= \text{sgn}\big(\mathbf{h}_{t,k}\big)^*\odot\text{sgn}\big(\mathbf{\Delta}^{(n_1)}_{\rho_{t,k}}\big)\odot\sqrt{\mathbf{p}_{t,k}^{(n_1)}},
\end{align}
where $\sqrt{\cdot}$ denotes the entry-wise square root function.

With the synchronized transmissions of $\mathbf{x}_{t,k}^{(n_1)}$ from all devices $k\in\mathcal{K}$ over the shared channels, the received signal \eqref{eq:ReceivedSignal} can be rewritten as
\begin{align}\label{eq:y_rho}
    \mathbf{y}_t^{(n_1)} &= \sum_{k\in\mathcal{K}}  \pi_k\sqrt{\frac{\gamma}{\bar{\delta}_{\rho_{t}}} } \text{sgn}\big(\mathbf{\Delta}_{\rho_{t,k}}^{(n_1)}\big)\odot\mathbf{v}_{\rho_{t,k}}^{(n_1)} + \mathbf{z}_t^{(n_1)},
\end{align}
where $\mathbf{z}_t^{(n_1)}\sim \mathcal{CN}(0, \sigma_z^2\mathbf{I}_F)$.
Then, the aggregated update vector is estimated as
\begin{align}\label{eq:delta_n}
    \hat{\mathbf{\Delta}}_{\rho_{t}}^{(n_1)} &= \sqrt{\frac{\bar{\delta}_{\rho_{t}}}{\gamma}} \mathbf{y}_t^{(n_1)} \nonumber\\
    &= \left(\sum_{k\in\mathcal{K}}\pi_k\mathbf{\Delta}_{\rho_{t,k}}^{(n_1)}\right)+\eta\boldsymbol{\xi}_{\rho_t}^{(n_1)} +\sqrt{\frac{\bar{\delta}_{\rho_{t}}}{\gamma}} \mathbf{z}_{t}^{(n_1)},
\end{align}
where $\boldsymbol{\xi}_{\rho_t}^{(n_1)}=\sum_{k\in\mathcal{K}}\pi_k\text{sgn}(\mathbf{\Delta}_{\rho_{t,k}}^{(n_1)})\odot\mathbf{e}_{\rho_{t,k}}^{(n_1)}$ denotes the aggregated update error, and $\mathbf{e}_{\rho_{t,k}}^{(n_1)}=\frac{1}{\eta}\big(\mathbf{v}_{\rho_{t,k}}^{(n_1)}-\big|\mathbf{\Delta}_{\rho_{t,k}}^{(n_1)}\big|\big)$ denotes the local update error resulting from power-constrained transmissions of device $k$.
The update in \eqref{eq:delta_n} can be interpreted as performing stochastic gradient updates with additive perturbations induced by wireless transmission.

By re-arranging the values of the vectors \eqref{eq:delta_n} for all $n\in\{1, 2, ..., N_1\}$ according to a predefined reverse grouping rule, the BS constitutes the aggregated composite update vector $\hat{\mathbf{\Delta}}_{\rho_t}\in\mathbb{R}^{d}$.
Based on the composite update vector, the BS updates the variational parameters associated with the covariance matrix of the global variational posterior as
\begin{align}\label{eq:rho_update}
    \boldsymbol{\rho}_{t+1} = \boldsymbol{\rho}_t + \hat{\mathbf{\Delta}}_{\rho_t}.
\end{align}
Then, the BS distributes the updated global covariance matrix $\mathbf{\Sigma}_{t+1}=\text{diag}(\boldsymbol{\rho}_{t+1})^{-1}$ to all devices.

In the second phase, for each device $k\in\mathcal{K}$, a new variational parameter vector $\boldsymbol{\nu}_{t,k}$ is defined using the received covariance matrix $\mathbf{\Sigma}_{t+1}$ and it is initialized as 
\begin{align}\label{eq:nu_def}
    \boldsymbol{\nu}_{t,k} \leftarrow \mathbf{\Sigma}_{t+1}\mathbf{\Sigma}_{t,k}^{-1}\boldsymbol{\mu}_t.
\end{align}
Using the updated global covariance as a common approximation, each device k generates an MC sample for local mean optimization as
\begin{align}
    \mathbf{w}_{(m)} = \mathbf{\Sigma}_{t+1}^{\frac{1}{2}}\boldsymbol{\epsilon}_{(m)} + \mathbf{\Sigma}_{t+1}^{-1} \mathbf{\Sigma}_{t,k} \boldsymbol{\nu}_{t,k}.
\end{align}
Similarly to \eqref{eq:update_phase1}, the local variational parameter vector $\boldsymbol{\nu}_{t,k}$ is updated at each gradient descent step as
\begin{align}
    \boldsymbol{\nu}_{t,k} \leftarrow \boldsymbol{\nu}_{t,k} -\eta \nabla_{\nu} L_k\!\left( \boldsymbol{\nu}_{t,k}, \text{diag}\big(\mathbf{\Sigma}_{t+1}^{-1}\big) \right).
\end{align}
After completing $E$ gradient descent steps, each device forms its local update vector $\mathbf{\Delta}_{\nu_{t,k}}=\boldsymbol{\nu}_{t,k}-\boldsymbol{\nu}_{t}$ and obtains the average update power $\bar{\delta}_{\nu_t}$ using the similar procedures described in \eqref{eq:rho_normal} and \eqref{eq:rho_avg}, respectively. 
It then partitions $\mathbf{\Delta}_{\nu_{t,k}}$ into $N_2$ groups and transmits the update values in each group via OFDM symbols $n_2\in\{N_1+1, N_1+2, ..., N_1+N_2\}$ according to the similar procedures described in \eqref{eq:PowerAlloc_1} and \eqref{eq:Gen_x}.
The BS estimates the aggregated update vector in a similar manner to \eqref{eq:y_rho}.
Then, the aggregated update vector for $\boldsymbol{\nu}_t$ is given by
\begin{align}\label{eq:delta_n2}
    \hat{\mathbf{\Delta}}_{\nu_{t}}^{(n_2)} &= \sqrt{\frac{\bar{\delta}_{\nu_{t}}}{\gamma}} \mathbf{y}_t^{(n_2)} \nonumber\\
    &= \left(\sum_{k\in\mathcal{K}}\pi_k\mathbf{\Delta}_{\nu_{t,k}}^{(n_2)}\right)+\eta\boldsymbol{\xi}_{\nu_t}^{(n_2)} +\sqrt{\frac{\bar{\delta}_{\nu_{t}}}{\gamma}} \mathbf{z}_{t}^{(n_2)},
\end{align}
where $\mathbf{z}_{t}^{(n_2)}\sim\mathcal{CN}(0, \sigma_z^2\mathbf{I}_F)$, $\boldsymbol{\xi}_{\nu_t}^{(n_2)}=\sum_{k\in\mathcal{K}}\pi_k\text{sgn}\big(\mathbf{\Delta}_{\nu_{t,k}}^{(n_2)}\big)\odot\mathbf{e}_{\nu_{t,k}}^{(n_2)}$, and $\mathbf{e}_{\nu_{t,k}}^{(n_2)}=\frac{1}{\eta}\big(\mathbf{v}_{\nu_{t,k}}^{(n_2)}-\big|\mathbf{\Delta}_{\nu_{t,k}}^{(n_2)}\big|\big)$.
The BS reconstructs the complete update vector $\hat{\mathbf{\Delta}}_{\nu_t}$ by re-arranging the update values delivered via $N_2$ OFDM symbols and updates the variational parameter vector as
\begin{align}
    \boldsymbol{\nu}_{t+1} = \boldsymbol{\nu}_t + \hat{\mathbf{\Delta}}_{\nu_t}.
\end{align}
Note that $\boldsymbol{\nu}_{t+1}$ is identical to $\boldsymbol{\mu}_{t+1}$ according to \eqref{eq:mu_sigma}.
The BS then proceeds to the next training round $t+1$ by distributing the updated global mean vector $\boldsymbol{\mu}_{t+1}$ to all devices if the convergence condition is not satisfied.

\begin{algorithm}[t!]
\caption{Bayesian FL with Over-the-Air Posterior Aggregation}
\label{Alg:FL_Wireless_BS}
\begin{algorithmic}[1]
\Initialize{$t\leftarrow 0$\\ $\boldsymbol{\mu}_t, \mathbf{\Sigma}_t \leftarrow$ Random initialization}

\State BS distributes $\boldsymbol{\mu}_0$ and $\mathbf{\Sigma}_0$ to all devices 
\While{convergence criterion is not met}
    \Statex \textit{// Phase 1: Update global covariance matrix}
    \For{each device $k\in\mathcal{K}$ \textbf{in parallel}}
        \State $\boldsymbol{\rho}_{t,k} \leftarrow \text{diag}\!\left( \mathbf{\Sigma}_{t}^{-1} \right)$
        \For {$i=1$ to $E$}
            \State $\boldsymbol{\rho}_{t,k}\leftarrow \boldsymbol{\rho}_{t,k} -\eta \nabla_\rho L_k(\boldsymbol{\mu}_t, \boldsymbol{\rho}_{t,k})$
        \EndFor
         \State Compute $\mathbf{\Delta}_{\rho_{t,k}}$ with \eqref{eq:rho_normal}
         \State Optimize power allocation $\{\mathbf{p}_{t,k}^{(n_1)}\}_{n_1=1}^{N_1}$
         \State Form $\big\{\mathbf{x}_{t,k}^{(n_1)}\big\}_{n_1=1}^{N_1}$ with \eqref{eq:Gen_x} and transmit them
    \EndFor
    \State BS reconstructs $\hat{\mathbf{\Delta}}_{\rho_t}$ from $\big\{\mathbf{y}_t^{(n_1)}\big\}_{n_1=1}^{N_1}$ with \eqref{eq:delta_n}
    \State $\boldsymbol{\rho}_{t+1}=\boldsymbol{\rho}_t+\hat{\mathbf{\Delta}}_{\rho_t}$
    \State BS distributes $\mathbf{\Sigma}_{t+1}=\text{diag}\big(\boldsymbol{\rho}_{t+1}\big)^{-1}$ to all devices
    \Statex \textit{// Phase 2: Update global mean vector}    
    \For{each device $k\in\mathcal{K}$ \textbf{in parallel}}
        \State $\boldsymbol{\nu}_{t,k} \leftarrow \mathbf{\Sigma}_{t,k}^{-1}\mathbf{\Sigma}_{t+1}\boldsymbol{\mu}_{t}$
        \For {$i=1$ to $E$}
            \State $\boldsymbol{\nu}_{t,k}\leftarrow \boldsymbol{\nu}_{t,k} -\eta \nabla_\nu L_k\Big(\boldsymbol{\nu}_{t,k}, \text{diag}\big(\mathbf{\Sigma}_{t+1}^{-1}\big)\Big)$
        \EndFor
         \State Compute the normalized update vector $\mathbf{\Delta}_{\nu_{t,k}}$ 
         \State Optimize power allocation $\big\{\mathbf{p}_{t,k}^{(n_2)}\big\}_{n_2=N_1+1}^{N_1+N_2}$
         \State Form $\big\{\mathbf{x}_{t,k}^{(n_2)}\big\}^{N_1+N_2}_{n_2=N_1+1}$  and transmit them
    \EndFor
    \State BS reconstructs $\hat{\mathbf{\Delta}}_{\nu_t}$ from $\big\{\mathbf{y}_t^{(n_2)}\big\}_{n_2=N_1+1}^{N_1+N_2}$ 
    \State $\boldsymbol{\nu}_{t+1}=\boldsymbol{\nu}_t+\hat{\mathbf{\Delta}}_{\nu_t}$
    \State BS distributes $\boldsymbol{\mu}_{t+1}=\boldsymbol{\nu}_{t+1}$ to all devices
    \State $t\leftarrow t+1$
\EndWhile
\end{algorithmic}
\end{algorithm}

\section{Convergence Analysis and Transmit Power Control}
In this section, we analyze the convergence of our proposed framework under given power allocations and investigate the impact of fading channels and limited power budget on the training convergence rate.
Subsequently, leveraging these observations, we optimize the transmit power allocation to enhance the training convergence. 

\subsection{Convergence Analysis}
For applications to deep neural networks, we investigate the convergence to stationary points, without assuming the convexity.

\begin{assumption}[Smooth] \label{ass:Smooth}
    For any $\boldsymbol{\nu}_1, \boldsymbol{\nu}_2, \boldsymbol{\rho}_1, \boldsymbol{\rho}_2 \in \mathbb{R}^{d}$, the global loss function $L(\boldsymbol{\nu}, \boldsymbol{\rho})$ is assumed to satisfy     
\begin{align}
    &|L(\boldsymbol{\nu}_1, \boldsymbol{\rho}) - L(\boldsymbol{\nu}_2, \boldsymbol{\rho}) - \nabla_\nu L(\boldsymbol{\nu}_2, \boldsymbol{\rho})^\intercal\!\left( \boldsymbol{\nu}_1-\boldsymbol{\nu}_2 \right)| \nonumber\\
    &\le  \frac{\Lambda_\nu}{2} \left\Vert \boldsymbol{\nu}_1-\boldsymbol{\nu}_2 \right\Vert^2
\end{align}
\begin{align}
    &|L(\boldsymbol{\nu}, \boldsymbol{\rho}_1) - L(\boldsymbol{\nu}, \boldsymbol{\rho}_2) - \nabla_\rho L(\boldsymbol{\nu}, \boldsymbol{\rho}_2)^\intercal\!\left( \boldsymbol{\rho}_1-\boldsymbol{\rho}_2 \right)|\nonumber\\
    &\le  \frac{\Lambda_\rho}{2} \left\Vert \boldsymbol{\rho}_1-\boldsymbol{\rho}_2 \right\Vert^2
\end{align}
for non-negative constants $\Lambda_\nu$ and $\Lambda_\rho$.
\end{assumption}

For simplicity, we assume $d=F$ and a single epoch $E=1$ for local training in the subsequent convergence analysis.
Consequently, the number of transmissions in each phase becomes $N_1=N_2=1$, and the local updates simplify to $\mathbf{\Delta}_{\rho_{t,k}}=-\eta\nabla_\rho L_k(\boldsymbol{\mu}_t, \boldsymbol{\rho}_t)$ and $\mathbf{\Delta}_{\nu_{t,k}}=-\eta\nabla_\nu L_k\big(\boldsymbol{\nu}_t, \text{diag}\big(\mathbf{\Sigma}_{t+1}^{-1}\big)\big)$.
\begin{theorem}\label{thr:CostReduction}
    Under Assumption \ref{ass:Smooth}, with the learning rate set as $\eta=\frac{1}{\sqrt{T}}$, the expected decrease in the cost function over $T$ training rounds is lower bounded by \eqref{eq:CostReduction}. 
    \begin{figure*}
        \begin{align}
    &\mathbb{E}\!\left[ L(\boldsymbol{\nu}_0, \boldsymbol{\rho}_0)-L(\boldsymbol{\nu}_T, \boldsymbol{\rho}_T) \right] \ge \frac{1}{2\sqrt{T}}\mathbb{E}\!\left[ \sum_{t=0}^{T-1}\left(1-\frac{\Lambda_\rho}{\sqrt{T}}\right) \left\Vert \nabla_\rho L\!\left(\boldsymbol{\mu}_t, \boldsymbol{\rho}_t\right)\right\Vert^2+\left(1-\frac{\Lambda_\nu}{\sqrt{T}}\right) \left\Vert \nabla_\nu L\!\left(\boldsymbol{\nu}_{t}, \text{diag}\!\left(\mathbf{\Sigma}_{t+1}^{-1}\right)\right)\right\Vert^2\right] \nonumber\\
    & - \underbrace{\frac{\sigma_z^2 }{2\gamma T} \mathbb{E}\!\left[\sum_{t=0}^{T-1} \sum_{k\in\mathcal{K}}\pi_k\left(\Lambda_\rho\left\Vert \nabla_\rho  L_k\!\left(\boldsymbol{\mu}_{t}, \boldsymbol{\rho}_{t,k}\right) \right\Vert^2+\Lambda_\nu\left\Vert \nabla_\nu  L_k\!\left(\boldsymbol{\nu}_{t,k}, \text{diag}\big(\mathbf{\Sigma}_{t+1}^{-1}\big) \right) \right\Vert^2\right) \right]}_{\text{Channel noise}} \nonumber\\
    &- \underbrace{\frac{1}{2\sqrt{T}}\mathbb{E}\!\left[\sum_{t=0}^{T-1}\left(1+\frac{\Lambda_\rho}{\sqrt{T}}\right)   \left\Vert\boldsymbol{\xi}_{\rho_t}\right\Vert^2 + \left(1 + \frac{\Lambda_\nu}{\sqrt{T}}\right) \left\Vert\boldsymbol{\xi}_{\nu_t}\right\Vert^2 \right]}_{\text{Limited transmission power budget}}
            \label{eq:CostReduction}
        \end{align}
        \hrulefill
    \end{figure*}
    \begin{proof}
        Refer to Appendix \ref{app:prrof_thm1}.
    \end{proof}
\end{theorem}

Theorem \ref{thr:CostReduction} shows that channel noise and update errors, $\boldsymbol{\xi}_{\rho_t}$ and $\boldsymbol{\xi}_{\nu_t}$, caused by the limited transmission power budget hinder the minimization of the cost function. 
Specifically, the second and third terms on the right hand side of \eqref{eq:CostReduction} explicitly capture the impacts of additive channel noise and update errors during wireless aggregation, respectively. Since both terms appear as non-positive penalties in the lower bound, they directly reduce the achievable decrease of the cost function, thereby slowing down the convergence of the training process under noisy communication conditions.
Note that update errors are highly dependent on channel conditions. Specifically, as shown in \eqref{eq:PowerAlloc_1}, when the entries of the channel power inversion vector $\mathbf{g}_{t,k}$ are large, the update magnitude vector, $\mathbf{v}_{\rho_{t,k}}$ or $\mathbf{v}_{\nu_{t,k}}$, must be reduced to meet the power constraint, leading to increased update errors.

\begin{corollary}\label{crl:AsymptoticBound}
    In the asymptotic regime as $T \to \infty$, the round-averaged squared $\ell_2$ norm of the global update vectors is upper-bounded by
    \begin{align}\label{eq:Asymp}
        &\mathbb{E}\bigg[\frac{1}{T} \sum_{t=0}^{T-1} \big\Vert \nabla_\rho L(\boldsymbol{\mu}_t, \boldsymbol{\rho}_t) \big\Vert^2 + \big\Vert \nabla_\nu L\big( \boldsymbol{\nu}_t, \text{diag}\big( \mathbf{\Sigma}_{t+1}^{-1}\big) \big) \big\Vert^2  \bigg] \nonumber\\
        &\le  \mathbb{E}\bigg[ \frac{1}{T}\sum_{t=0}^{T-1}\Vert \boldsymbol{\xi}_{\rho_t} \Vert^2 \! +\! \Vert \boldsymbol{\xi}_{\nu_t} \Vert^2\bigg] +\frac{1}{\sqrt{T}}\big( L(\boldsymbol{\nu}_0, \boldsymbol{\rho}_0)\!-\!L(\boldsymbol{\nu}^*, \boldsymbol{\rho}^*) \big),
    \end{align}
    where $\boldsymbol{\nu}^*$ and $\boldsymbol{\rho}^*$ denote the optimal variational parameter vectors that minimize the global cost function.
    \begin{proof}
        In the asymptotic regime as $T \to \infty$, combining Theorem 1 with the inequality $L(\boldsymbol{\nu}_T, \boldsymbol{\rho}_T) \ge L(\boldsymbol{\nu}^*, \boldsymbol{\rho}^*)$ yields \eqref{eq:Asymp}.
    \end{proof}
\end{corollary}

Corollary \ref{crl:AsymptoticBound} indicates that the influence of communication noise on convergence becomes negligible after a sufficiently large number of training rounds due to the averaging effect across successive updates, which is consistent with the convergence behavior observed in conventional AirComp-based FL schemes \cite{JPHong2023_OTA_FL, Cui2022_AirComp}.
Consequently, as $T \to \infty$, the iterates do not necessarily converge to a fixed point but instead fluctuate around a stationary point, in the sense that the averaged squared gradient norm is upper-bounded by the round-averaged aggregated distortion power. If the round-averaged aggregated distortion power vanishes as $T \to \infty$, exact first-order stationarity is recovered.
Interestingly, although the proposed framework adopts a distinct approach from conventional AirComp-based FL \cite{Cui2022_AirComp, ICC2023, JPHong2023_OTA_FL} in terms of local training and update aggregation, it exhibits similar convergence behavior, in that the learning performance is largely influenced by the aggregated update error.

\subsection{Transmit Power Optimization}
From Theorem 1 and Corollary 1, the convergence behavior of the proposed framework is primarily determined by the aggregated update distortion introduced during wireless transmission. 
Therefore, the objective of the power control strategy is to reduce this distortion term, thereby directly accelerating the convergence of the learning process.

In the proposed framework, the transmit signal is governed by the update magnitude vector, $\mathbf{v}_{\rho_{t,k}}^{(n_1)}$ or $\mathbf{v}_{\nu_{t,k}}^{(n_2)}$, as defined in \eqref{eq:PowerAlloc_1} and \eqref{eq:Gen_x}. 
Given that the aggregated update errors from the two variational parameters have an equal impact on training convergence as shown in \eqref{eq:Asymp}, the same power optimization process is applied to both $\mathbf{v}_{\rho_{t,k}}^{(n_1)}$ and $\mathbf{v}_{\nu_{t,k}}^{(n_2)}$.  
To avoid redundancy, we focus on the optimization of $\mathbf{v}_{\rho_{t,k}}^{(n_1)}$ in the remainder of this subsection.
To enable distributed power control, we minimize a separable upper bound on the aggregated update distortion. 
Specifically, since $\sum_{k\in\mathcal{K}}\pi_k=1$, Jensen's inequality yields
\begin{align}
    \left\|\boldsymbol{\xi}_{\rho_t}^{(n_1)}\right\|_2^2 &= \left\|\sum_{k\in\mathcal{K}}\pi_k\operatorname{sgn}\!\left(\mathbf{\Delta}_{\rho_{t,k}}^{(n_1)}
    \right)\odot\mathbf{e}_{\rho_{t,k}}^{(n_1)}\right\|_2^2 \nonumber\\
    &\leq\sum_{k\in\mathcal{K}}\pi_k \left\| \operatorname{sgn}\!\left( \mathbf{\Delta}_{\rho_{t,k}}^{(n_1)} \right) \odot \mathbf{e}_{\rho_{t,k}}^{(n_1)}
    \right\|_2^2 \nonumber\\
    &\leq \sum_{k\in\mathcal{K}}\pi_k\left\| \mathbf{e}_{\rho_{t,k}}^{(n_1)} \right\|_2^2.    \label{eq:distortion_upper_bound}
\end{align}
Because the transmit power constraints are separable across devices, minimizing the local update error at each device jointly minimizes the right-hand side of
\eqref{eq:distortion_upper_bound}. 
Accordingly, device $k$ solves the following optimization problem:
\begin{align}\label{eq:OptPow}
    &\underset{\mathbf{v}\in\mathbb{R}^{F}}{\text{minimize}} ~ \big\Vert \big\vert\mathbf{\Delta}_{\rho_{t,k}}^{(n_1)}\big\vert- \mathbf{v}\big\Vert^2 \\
    &\text{subject to } ~~\mathbf{u}_{t,k}^\intercal ( \mathbf{v} \odot \mathbf{v} ) \le P \nonumber\\
    &\qquad \qquad ~~~~ v_f\ge 0, ~~~\qquad f=1, 2, ..., F \nonumber
\end{align}

The power optimization \eqref{eq:OptPow} is a quadratically constrained quadratic programming (QCQP) problem, and its solution, derived from the Karush-Kuhn-Tucker (KKT) conditions, is given by
\begin{align}\label{eq:Opt_v}
    &\mathbf{v}_{\rho_{t,k}}^{(n_1)} \nonumber\\
    &= \left\{  \begin{array}{ll}
        \!\!\!\vert \mathbf{\Delta}_{\rho_{t,k}}^{(n_1)} \vert, & \text{if } \mathbf{u}_{t,k}^\intercal \big( \mathbf{\Delta}_{\rho_{t,k}}^{(n_1)}\!\odot\! \mathbf{\Delta}_{\rho_{t,k}}^{(n_1)}\big)\!\le\! P\\
        \!\!\!\vert \mathbf{\Delta}_{\rho_{t,k}}^{(n_1)} \vert\!\oslash\!\big(\!\boldsymbol{1}_F\!+\!\lambda_{\rho_{t,k}}^{(n_1)} \mathbf{u}_{t,k}\! \big) & \text{Otherwise } 
    \end{array},
    \right.
\end{align}
where the operation $\oslash$ denotes the Hadamard (element-wise) division, $\boldsymbol{1}_F\in\mathbb{R}^F$ is the all-ones vector, and $\lambda_{\rho_{t,k}}^{(n_1)}$ is a scalar satisfying $\mathbf{u}_{\rho_{t,k}}^\intercal\big( \mathbf{v}_{\rho_{t,k}}^{(n_1)} \odot \mathbf{v}_{\rho_{t,k}}^{(n_1)} \big)=P$.
From the expression of the optimized power control vector in \eqref{eq:Opt_v}, we observe that device $k$ can transmit its local update without reducing the update magnitudes when the updates are small or the elements of $\mathbf{u}_{\rho_{t,k}}$ are low. The latter typically occurs under favorable channel conditions or with a small local dataset. Otherwise, the update magnitude must be scaled down to satisfy the transmit power constraint. In particular, the reduction factor $\big(1+\lambda_{\rho_{t,k}}^{(n_1)}u_{t,k,f}\big)$ is inversely proportional to the channel power gain of each sub-channel. For example, a local update value transmitted over sub-channel $f$ must be significantly attenuated if the channel gain of that sub-channel is poor.

It is worth noting that the optimized power control in \eqref{eq:Opt_v}, along with the associated discussion, is also applicable to the AirComp-based aggregation in the second phase by substituting $\mathbf{\Delta}_{\rho_{t,k}}^{(n_1)}$ with $\mathbf{\Delta}_{\nu_{t,k}}^{(n_2)}$.

\section{Simulation Results}
All simulation codes were implemented in Python 3.12.7 using NumPy 1.26.4, PyTorch 2.5.1, bayesian-torch 0.5.0, and the Blitz package \cite{esposito2020blitzbdl}.
As benchmark methods, we consider several representative FL algorithms designed to address data heterogeneity: FedAvg \cite{McMahan2017}, FedProx \cite{Smith2020_FedProx}, SCAFFOLD \cite{Theertha2020_SCAFFOLD}, FedSMOO \cite{FedSMOO}, and FedGMT-v2 \cite{li2025fedgmt}. These methods employ a frequentist approach for local training, and we apply the optimized power allocation method developed in \cite{JPHong2023_OTA_FL} for amplitude alignment over wireless networks.

\begin{table}
    \centering
        \caption{Simulation Environment}

    \begin{tabular}{c|l|l}
        \hline
        \textbf{Symbol} & \textbf{Description} & \textbf{Value} \\ \hline\hline 
        $r_\text{cvge}$ & BS coverage radius  & $300$ m \\ \hline
        $K$ & No. of edge devices  & $100$ \\ \hline
        $|\mathcal{D}_k|$ & \makecell[l]{Local dataset size \\ (Poisson r.v.)}  & \makecell[l]{MNIST: mean $10$\\
                      CIFAR-100: mean $100$} \\ \hline
        $\{\mathcal{D}_k\}_{k\in\mathcal{K}}$ & Local data distribution & \makecell[l]{MNIST: single-class-per-device\\ 
        CIFAR-100: Dirichlet $\alpha=0.1$} \\ \hline
        $P$ & \makecell[l]{Maximum OFDM \\ symbols power}  & 20 dBm   \\ \hline
        $\sigma_z^2$ & Effective noise power & -74 dBm    \\ \hline
        $F$ & \makecell[l]{Number of OFDM \\ sub-channels} & $1{,}024$ \\ \hline
        $N_1, N_2$ & \makecell[l]{Number of OFDM \\ symbols per phase}  & \makecell[l]{MNIST: 456\\ 
        CIFAR-100: 838} \\ \hline
        $\alpha_\text{PL}$ & Path loss exponent  & $4$ \\ \hline
    \end{tabular}
    \label{table:SimulSetup}
\end{table}
\begin{table}
    \centering
    \caption{Training Hyper-Parameters}
    \begin{tabular}{c|l|l}
        \hline
        \textbf{Symbol} & \textbf{Description} & \textbf{Value} \\ \hline\hline
        $\eta$ & Learning rate in local training & \makecell[l]{MNIST: 0.1 \\ CIFAR-100: 0.05} \\ \hline
        $\lambda$ & KL regularization scaling factor &  $K$\\ \hline
        $B$ & Mini-batch size in local training & $|\mathcal{D}_k|$ \\ \hline
        $E$ & Local training epoch & \makecell[l]{MNIST: 3 \\ CIFAR-100: 30} \\ \hline
        $d$ & Number of DNN weights & \makecell[l]{MNIST: 466,698 \\ CIFAR-100: 857,736}\\ \hline
        $\gamma$ & Coefficient for power scaling & $\sigma_z^2$ \\ \hline
        $M$ & Number of MC samplings & \makecell[l]{MNIST: 5 \\ CIFAR-100:1} \\ \hline
    \end{tabular}
    \label{table:HyperParam}
\end{table}

We consider a wireless edge network scenario in which 100 edge devices are uniformly distributed within a radius of 300 m centered at the BS. 
The MNIST and CIFAR-100 datasets are used for the simulations.
For MNIST, each device is assigned $\ell$ labels, and each local sample is independently drawn from the assigned label set with equal probability $1/\ell$. 
The local dataset size of each device is drawn from a Poisson distribution. 
Unless otherwise specified, we set $\ell=1$, corresponding to the single-class-per-device setting. 
This deliberately severe setting models data-scarce devices that collect only a handful of observations from a single localized source or event category, for which a single-class local dataset is plausible.
For CIFAR-100, training samples are distributed across 100 devices using a device-wise Dirichlet--multinomial partition.
Specifically, the local dataset size of each device $k$ is drawn from a shifted Poisson distribution with a mean of 100 samples.
The class-proportion vector of device $k$ is independently drawn as $\boldsymbol{p}_k \sim \operatorname{Dirichlet}(\alpha\mathbf{1}_C)$, where $C=100$. The labels of its local samples are then independently drawn according to $\boldsymbol{p}_k$, and the corresponding images are sampled without replacement from the class-specific data pools.
Unless otherwise specified, we set $\alpha=0.1$, under which each device contains samples from an average of 21.6 out of the 100 classes. This setting allows the proposed framework to be evaluated under a commonly adopted Dirichlet-based statistical heterogeneity model while retaining the data-scarce nature of the considered wireless edge-learning scenario.

Unless otherwise indicated, all experimental results are obtained using the simulation setup and training hyper-parameters summarized in Tables \ref{table:SimulSetup} and \ref{table:HyperParam}, respectively.
Note that the limited and skewed training data in this scenario makes the learning performance highly sensitive to the randomness in both the local dataset size and the assigned label for each device. 
As a result, even with identical simulation settings described in Tables \ref{table:SimulSetup} and \ref{table:HyperParam}, the outcomes can vary considerably depending on how data is distributed across devices. To account for this effect and ensure a fair performance evaluation, all simulation results are averaged over 5 independent realizations of local dataset sizes and class assignments.

\begin{figure*}[t]
    \centering
    \subfigure[MNIST]{%
        \includegraphics[width=0.45\textwidth]{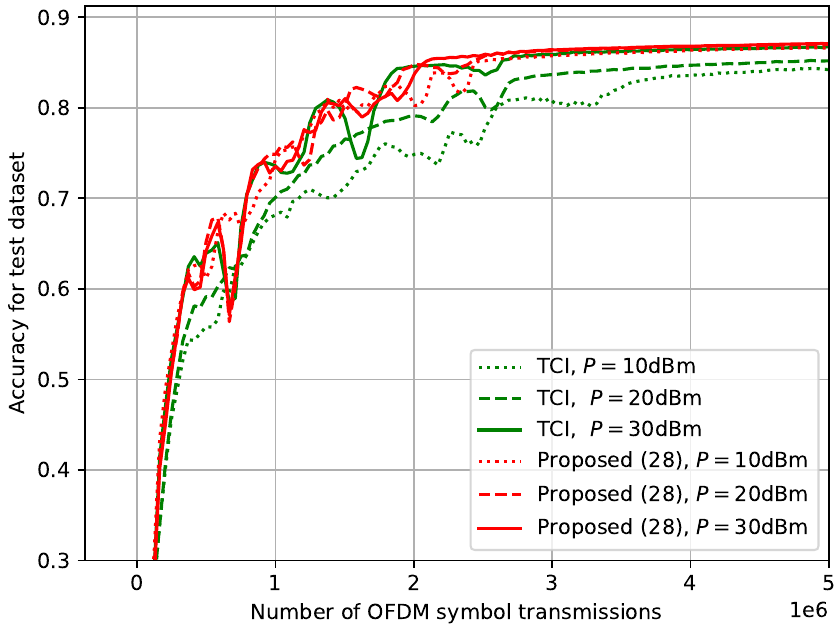}
        \label{fig:Pow_Alloc_MNIST}
    }
    \hfill
    \subfigure[CIFAR-100]{%
        \includegraphics[width=0.46\textwidth]{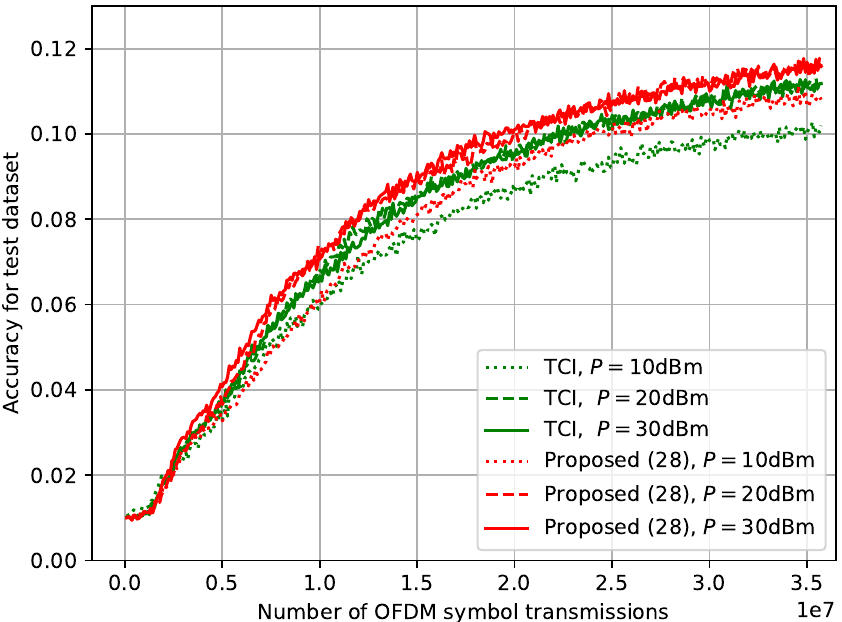}
        \label{fig:Pow_Alloc_CIFAR100}
    }
    \caption{Comparison of power allocation strategies for AirComp.}
    \label{fig:Pow_Alloc}
\end{figure*}

\begin{figure*}[t]
    \centering
    \subfigure[MNIST]{%
        \includegraphics[width=0.45\textwidth]{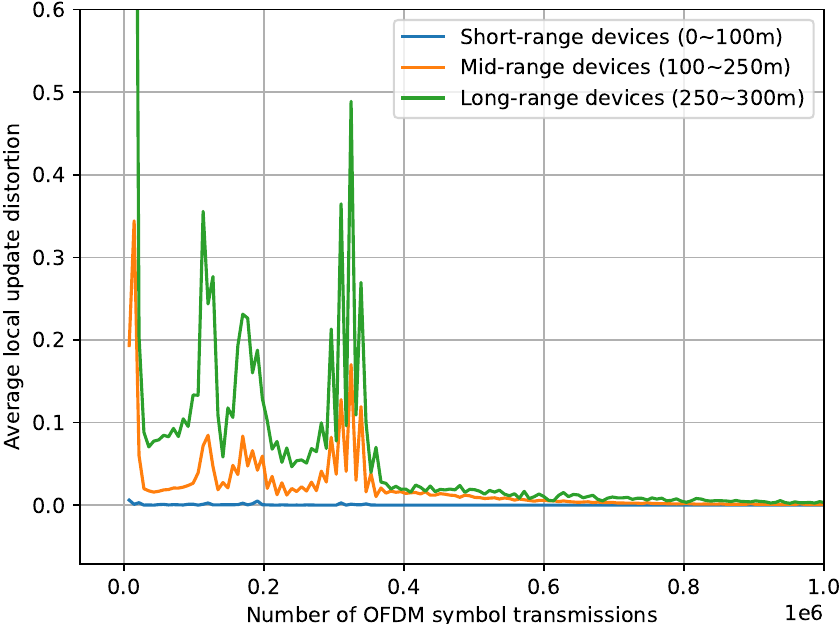}
        \label{fig:Distortion_test_MNIST}
    }
    \hfill
    \subfigure[CIFAR-100]{%
        \includegraphics[width=0.46\textwidth]{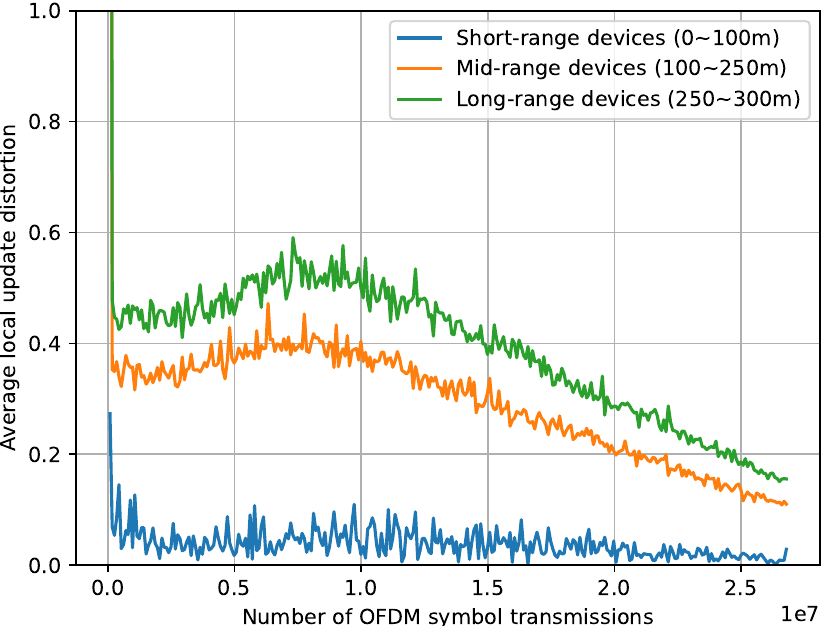}
        \label{fig:Distortion_test_CIFAR100}
    }
    \caption{Average local update distortion across devices in different distance ranges.}
    \label{fig:distortion_test}
\end{figure*}

Through FL, we train dataset-specific neural network architectures. 
For MNIST, we employ a six-layer multi-layer perceptron (MLP) with 256 hidden units per layer and ReLU activation functions, comprising 
$d=466{,}698$ model weights.
For CIFAR-100, we employ ResNet-56 to evaluate the proposed framework on a more complex image-classification task using a deep residual architecture. The corresponding number of model weights is $d=857{,}736$.
FedAvg \cite{McMahan2017}, FedProx \cite{Smith2020_FedProx}, and FedGMT-v2 \cite{li2025fedgmt} transmit $d$ real values in each of the downlink and uplink communications per training round. In contrast, SCAFFOLD \cite{Theertha2020_SCAFFOLD}, FedSMOO \cite{FedSMOO}, and the proposed framework transmit $2d$ real values in each communication direction per round. For SCAFFOLD, these consist of $d$ model parameters and $d$ control-variate parameters used to mitigate client drift. For FedSMOO, the additional $d$ values correspond to auxiliary variables introduced to stabilize local training under data heterogeneity. 
FedGMT-v2 avoids transmitting an additional model-sized variable by locally approximating the global model trajectory. 
For the proposed framework, the $2d$ values represent the mean and variance parameters of the posterior distributions over the model weights.

\begin{figure*}[t]
    \centering
    \subfigure[MNIST]{%
        \includegraphics[width=0.46\textwidth]{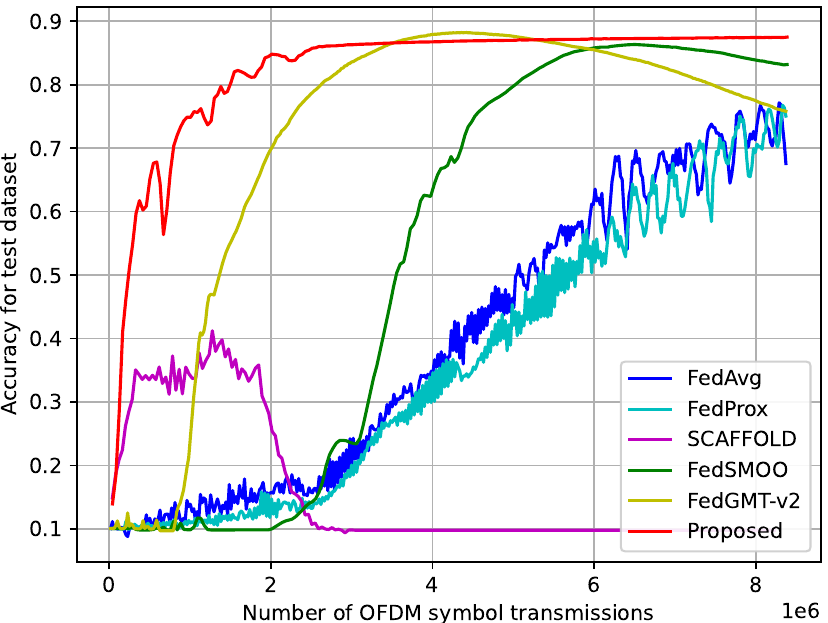}
        \label{fig:Algs_MNIST}
    }
    \hfill
    \subfigure[CIFAR-100]{%
        \includegraphics[width=0.47\textwidth]{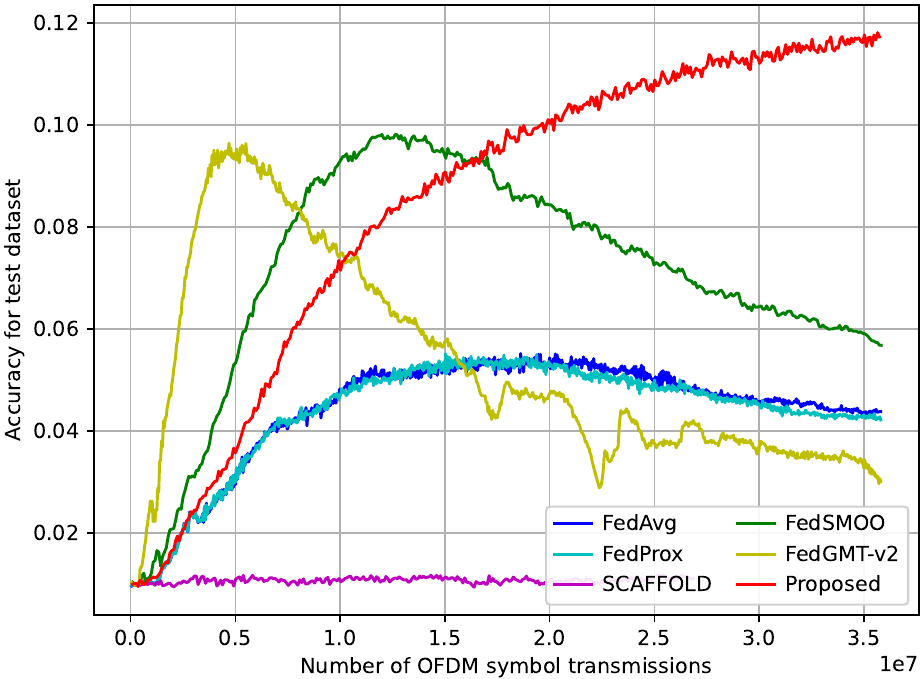}
        \label{fig:Algs_CIFAR100}
    }
    \caption{Comparison of wireless FL methods in scenario with scarce and heterogeneous data}\label{fig:Algs}
\end{figure*}

\begin{figure*}[t]
    \centering
    \subfigure[MNIST]{%
        \includegraphics[width=0.46\textwidth]{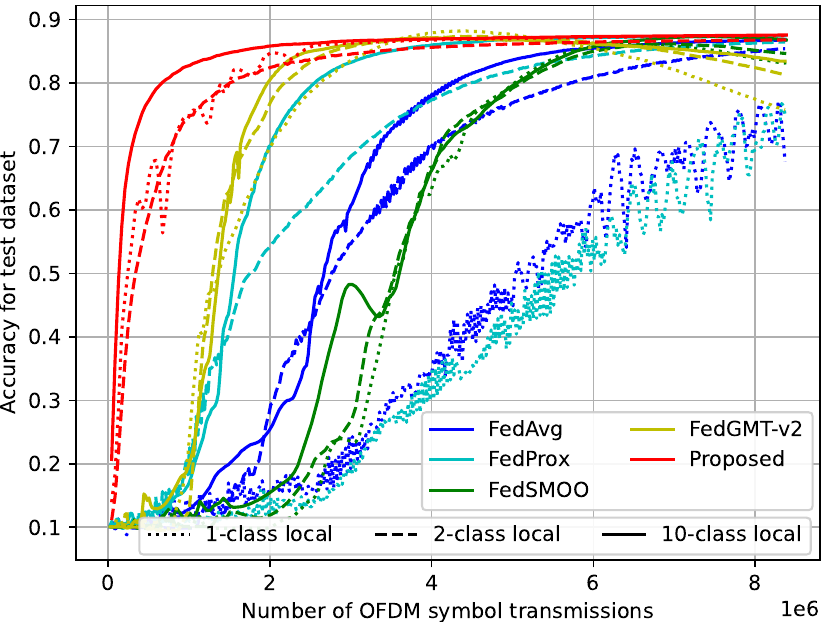}
    }
    \hfill
    \subfigure[CIFAR-100]{%
        \includegraphics[width=0.47\textwidth]{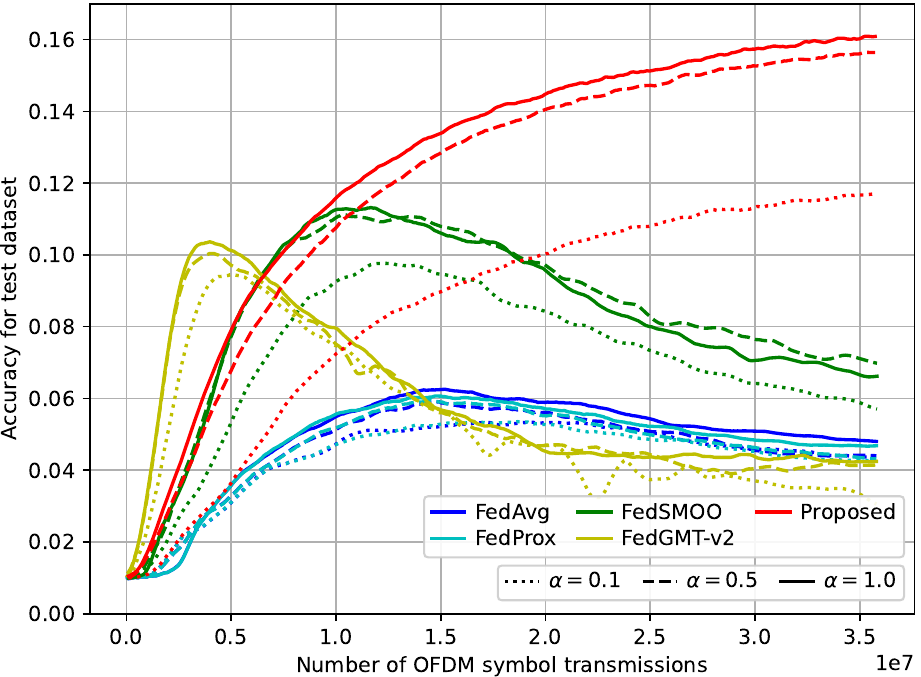}
    }
    \caption{Training progress under different degrees of local label skewness}\label{fig:VaryingLabels}
\end{figure*}

\subsection{Analysis of the Proposed Framework}
This subsection examines the behavior of the proposed framework under the considered wireless FL setting. To this end, we present a set of targeted simulations that provide insight into its operational characteristics and update aggregation behavior.

First, to evaluate the effectiveness of the proposed power allocation strategy based on (28) and (43) in facilitating training convergence, figure \ref{fig:Pow_Alloc} compares the proposed framework with a variant employing the truncated channel inversion (TCI) power allocation strategy, a widely used approach for AirComp in the literature \cite{Gundz2020_AirComp, Huang2020_AirComp, Poor2021_TCI, Eldar2024_TCI}, for the MNIST and CIFAR-100 datasets.
The proposed power control strategy is shown to achieve faster convergence and improved final performance compared to the TCI baseline under a limited power budget across both datasets. This performance gain stems from the fact that the proposed scheme is analytically designed to minimize a separable upper bound on the aggregation distortion, thereby accelerating convergence under power constrained transmission. In contrast, the TCI strategy primarily focuses on channel equalization without explicitly accounting for its impact on learning convergence.
These observations are consistent with the theoretical analysis in Section V, which indicates that communication-induced aggregation distortion becomes the dominant factor limiting convergence under constrained transmit power. Therefore, directly controlling this distortion through optimized power allocation is critical for improving learning performance. By jointly considering channel conditions and learning dynamics, the proposed power control strategy enhances aggregation reliability and improves overall learning performance in communication-constrained wireless environments.
Furthermore, the performance gap between the proposed strategy and the TCI baseline diminishes as the transmit power budget increases. This is because, under sufficiently large transmit power, both strategies approach full channel inversion, effectively mitigating channel fading and reducing aggregation distortion.

Figure~\ref{fig:distortion_test} illustrates the evolution of the average local update distortion across devices in short-, mid-, and long-range regions as a function of the number of OFDM symbol transmissions. 
As expected, devices located farther from the BS experience larger distortion due to stronger path loss and smaller channel gains, resulting in more frequent attenuation of their transmitted updates. 
In particular, mid-range and long-range devices exhibit noticeable distortion in the initial training rounds due to relatively severe channel conditions.
Nevertheless, the distortion gradually decreases as the training progresses and approaches zero for all device groups. 
This reduction can be attributed to the fact that, as training proceeds, the magnitude of local updates decreases and the variation of updates across devices becomes smaller, leading to reduced aggregation mismatch and attenuation effects. 
This behavior is consistent with the convergence analysis, which indicates that the learning process converges when the aggregated distortion diminishes over successive training rounds. 
The result demonstrates that, despite the presence of attenuation caused by deep fading and limited transmit power, the proposed framework effectively mitigates the impact of distortion and maintains stable convergence under the considered wireless environment.

\subsection{Convergence and Accuracy Performances}
In this subsection, we present simulation results on the test accuracy of FL methods as a function of the number of channel uses for downlink multicast and uplink local update aggregation under diverse environments.
These results allow us to assess both the convergence speed and the test accuracy upon convergence, as well as to understand the effect of system parameters on training performance.

\begin{figure*}[t]
    \centering
    \subfigure[MNIST]{%
        \includegraphics[width=0.46\textwidth]{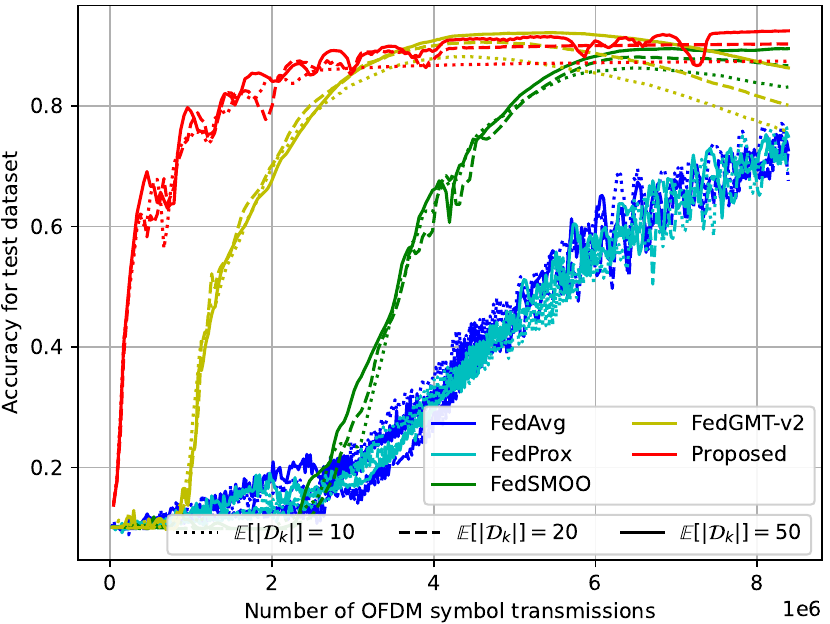}
    }
    \hfill
    \subfigure[CIFAR-100]{%
        \includegraphics[width=0.47\textwidth]{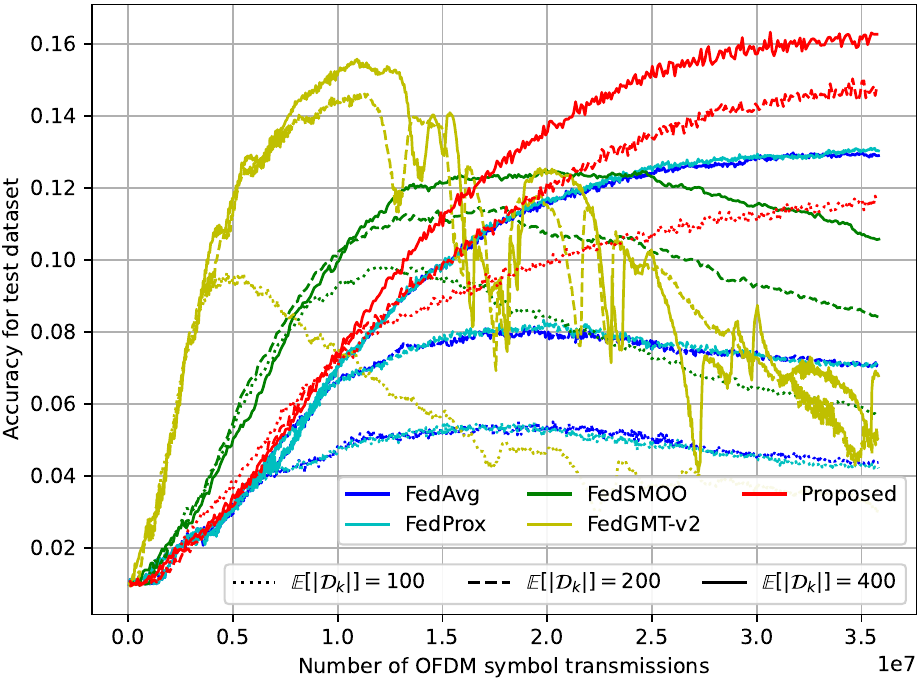}
    }
    \caption{Training progress for different local dataset sizes}\label{fig:VaryingSizes}
\end{figure*}

\begin{figure*}[t]
    \centering
    \subfigure[MNIST]{%
        \includegraphics[width=0.46\textwidth]{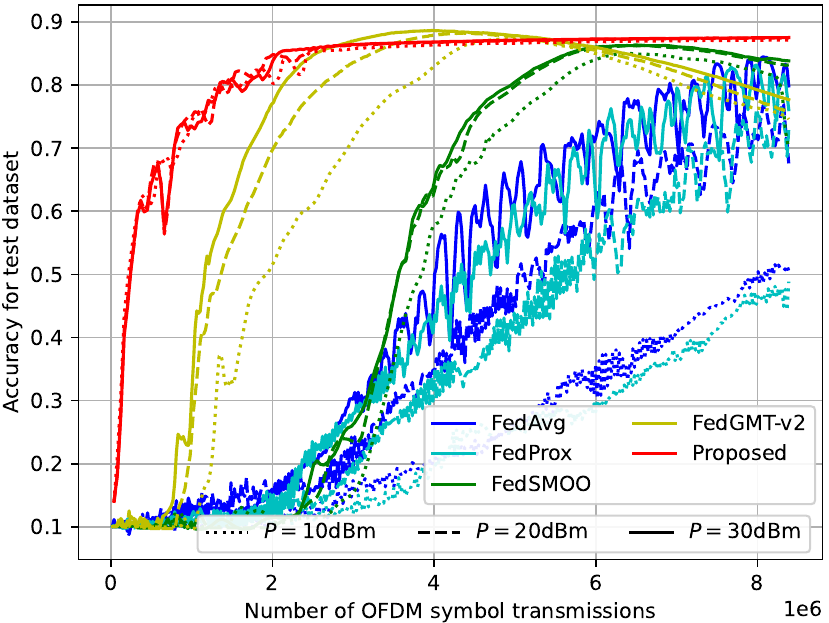}
    }
    \hfill
    \subfigure[CIFAR-100]{%
        \includegraphics[width=0.47\textwidth]{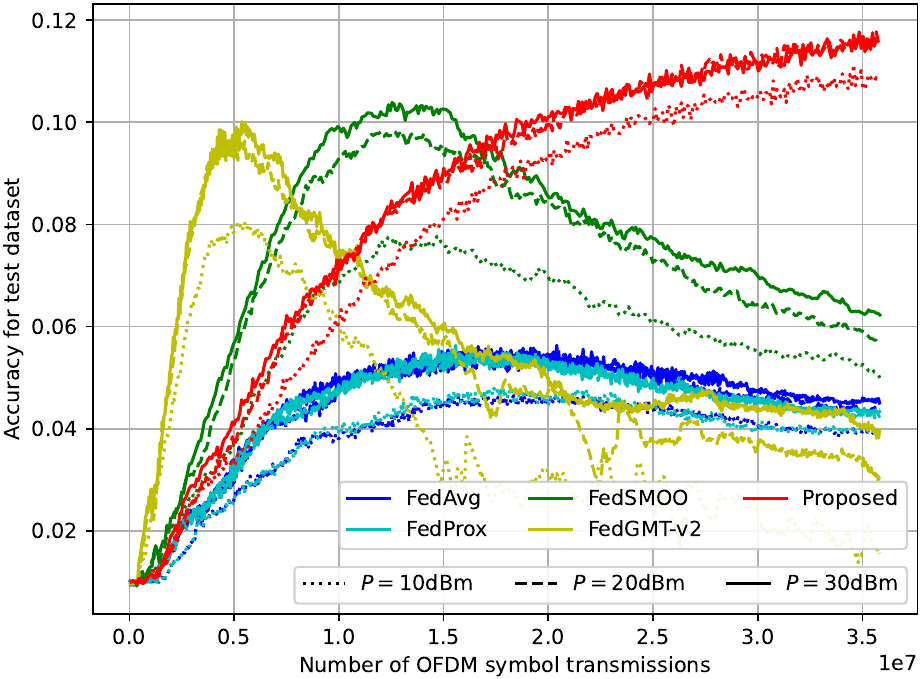}
    }
    \caption{Training progress for different transmission power budgets}\label{fig:PowerBudget}
\end{figure*}

Figure \ref{fig:Algs} presents the training progress of the proposed framework and benchmark FL methods as a function of the number of channel uses required for update aggregation.
The proposed framework demonstrates higher test accuracy compared to the benchmark methods. 
This improvement is attributed to Bayesian local training and distribution-level aggregation, which mitigate local overfitting and client drift while accounting for posterior uncertainty. Although the proposed framework communicates two sets of posterior parameters and therefore requires approximately twice the per-round communication payload of FedAvg and FedProx, this additional overhead does not result in a corresponding convergence delay. The resulting improvement in learning progress per round largely compensates for the increased communication cost. FedSMOO and FedGMT-v2 also exhibit rapid initial convergence and attain relatively high peak accuracies; however, their performance deteriorates as training continues, whereas the proposed framework maintains its attained accuracy. FedGMT-v2 reaches its peak after approximately half the number of OFDM symbol transmissions required by FedSMOO because of its lower per-round communication overhead, but exhibits a similar post-peak degradation pattern. This behavior is consistent with the sensitivity of SAM-based correction mechanisms to high-variance local updates and transmission-induced perturbations.
Furthermore, SCAFFOLD is prone to collapse due to its high sensitivity to noise in the control variate. For this reason, it is not well-suited to analog AirComp environments where noisy signal recovery is unavoidable; therefore, SCAFFOLD is excluded from the following simulation results.

Figure~\ref{fig:VaryingLabels} illustrates the training progress of the proposed framework and benchmark methods under varying degrees of local label skewness. 
On MNIST, all methods achieve similarly high final test accuracy when each device has samples from all 10 classes. As the number of local classes decreases, however, FedAvg and FedProx suffer substantial performance degradation, whereas the proposed framework and the SAM-based methods remain comparatively robust. On CIFAR-100, the proposed framework exhibits stable learning performance across all considered values of $\alpha$. Although it requires more OFDM symbol transmissions than the SAM-based methods, FedSMOO and FedGMT-v2, to reach its peak accuracy, it maintains the attained performance as training proceeds, whereas the SAM-based methods exhibit post-peak degradation. These results demonstrate the robustness of the proposed framework to heterogeneous local label distributions.

\renewcommand{\thesubfigure}{} 

\begin{figure*}[t]
    \centering

    \subfigure[FedAvg]{%
        \includegraphics[width=0.185\textwidth]{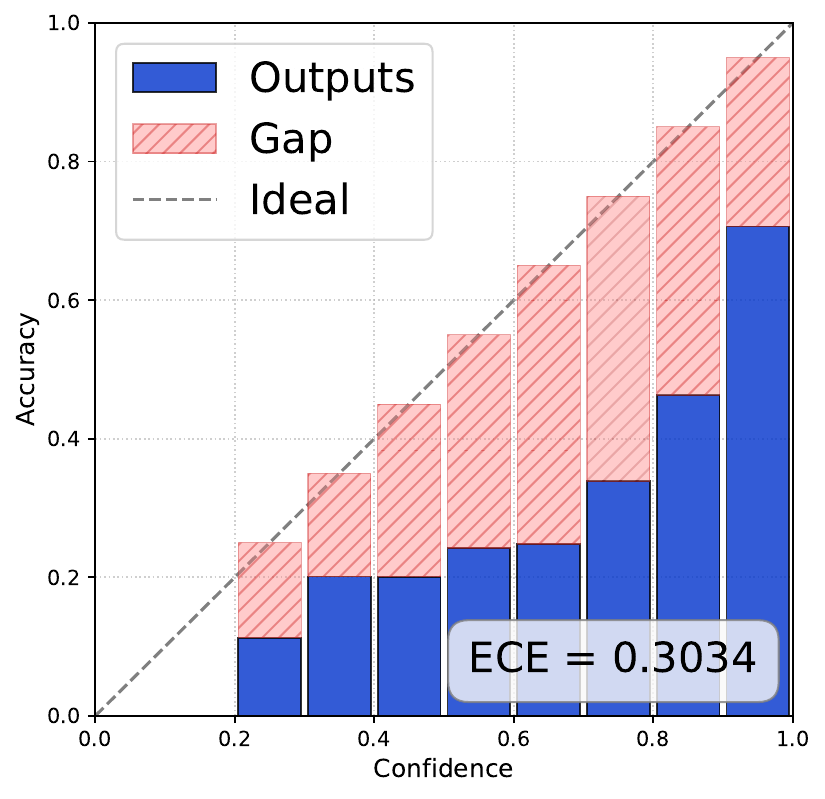}
    }
    \subfigure[FedProx]{%
        \includegraphics[width=0.185\textwidth]{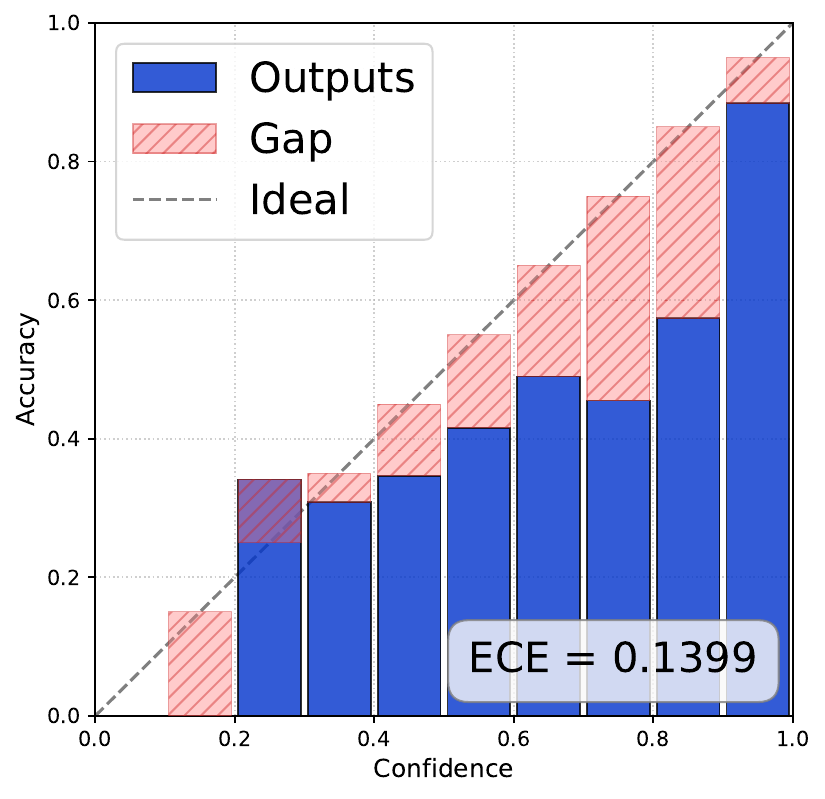}
    }
    \subfigure[FedSMOO]{%
        \includegraphics[width=0.185\textwidth]{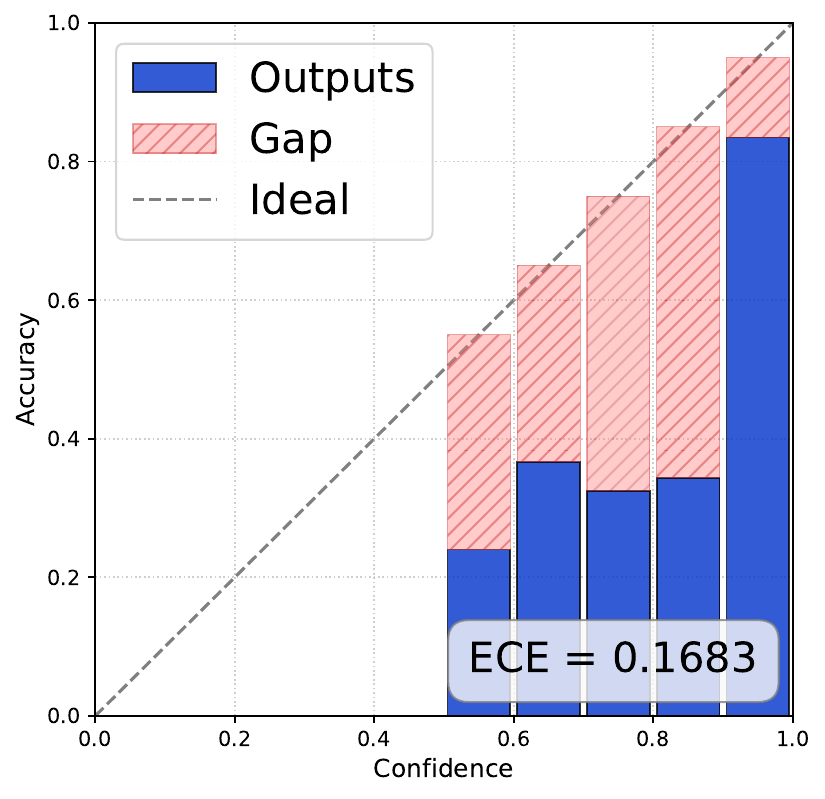}
    }
    \subfigure[FedGMT-v2]{%
        \includegraphics[width=0.185\textwidth]{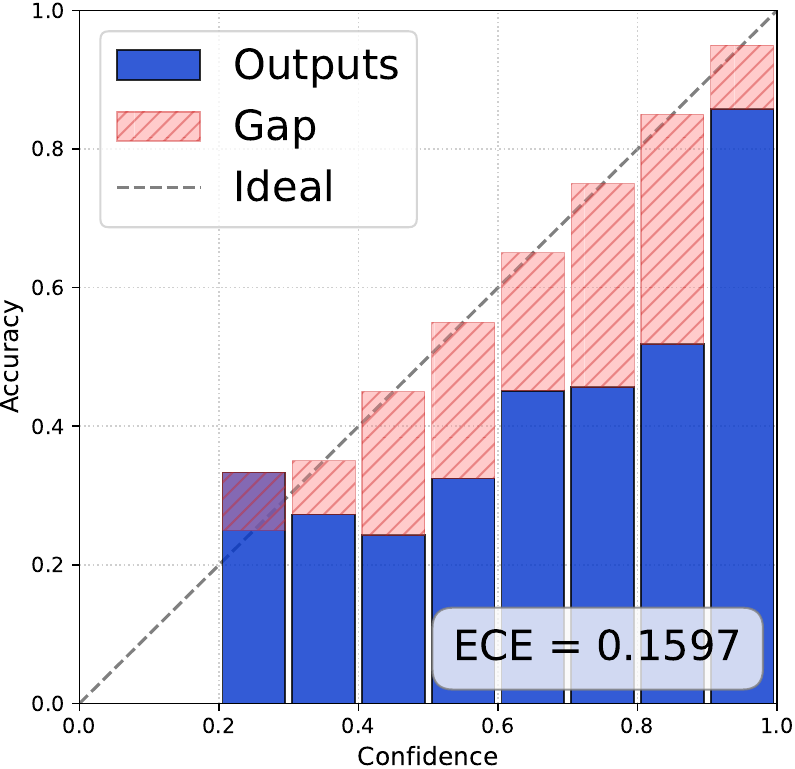}
    }
    \subfigure[Proposed]{%
        \includegraphics[width=0.185\textwidth]{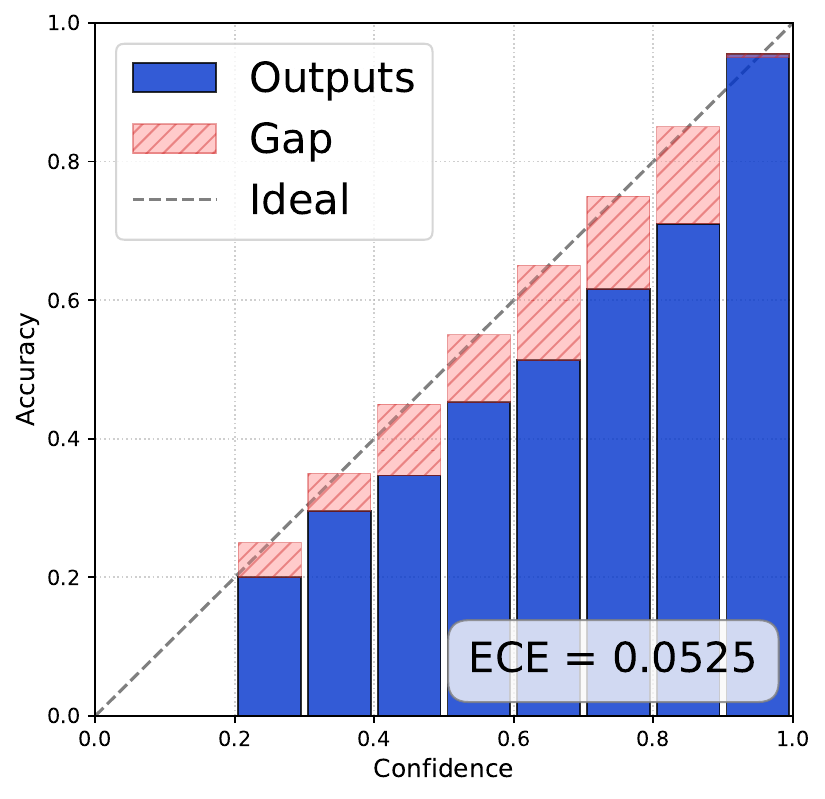}
    }

    {\centering \small (a) MNIST\par}

    \vspace{0.3cm}

    \subfigure[FedAvg]{%
        \includegraphics[width=0.185\textwidth]{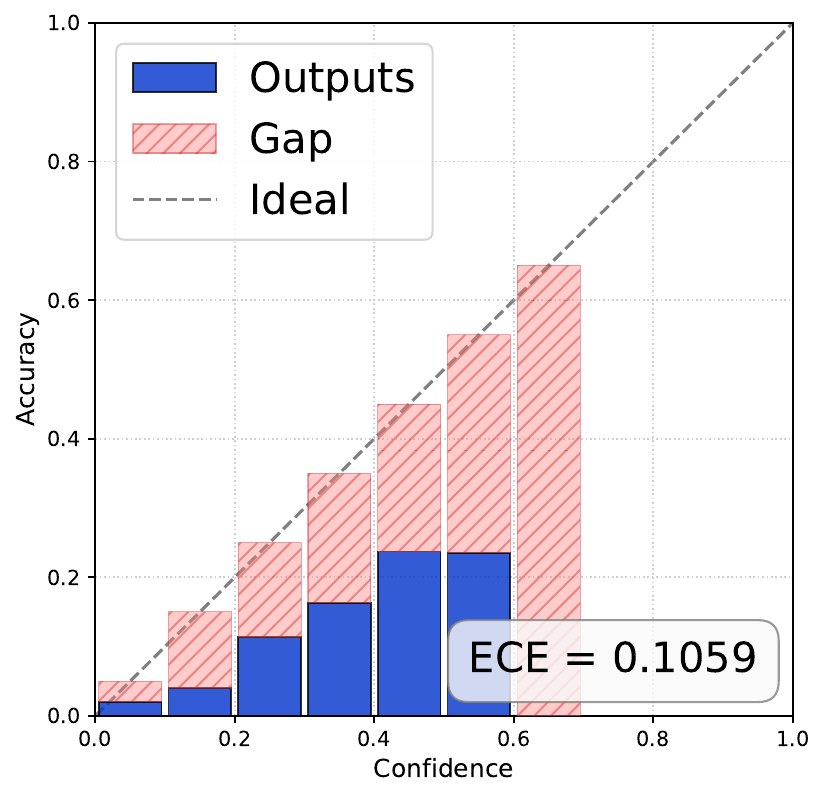}
    }
    \subfigure[FedProx]{%
        \includegraphics[width=0.185\textwidth]{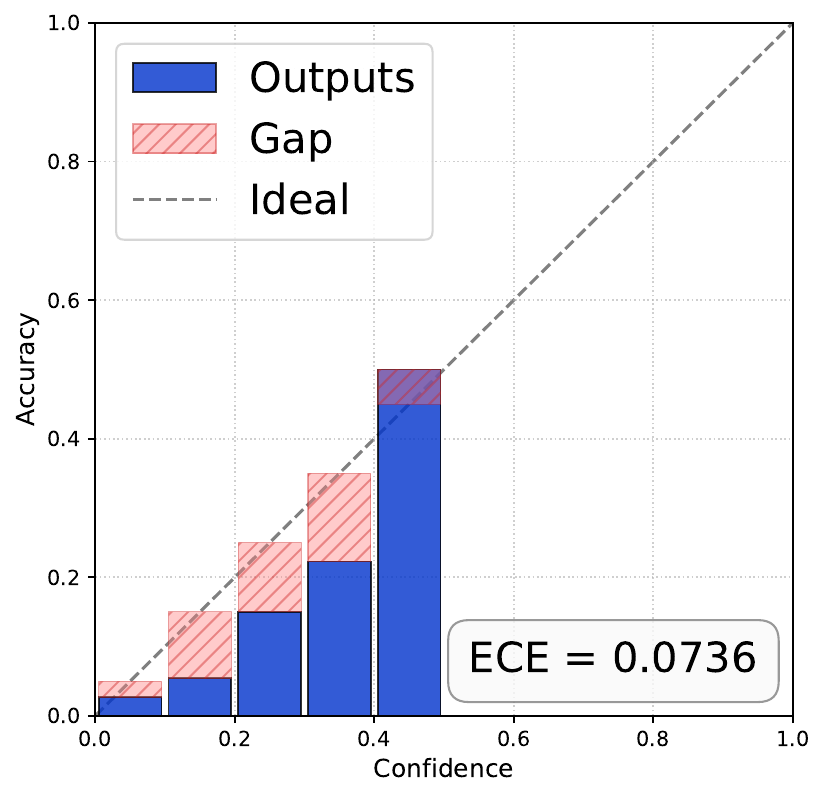}
    }
    \subfigure[FedSMOO]{%
        \includegraphics[width=0.185\textwidth]{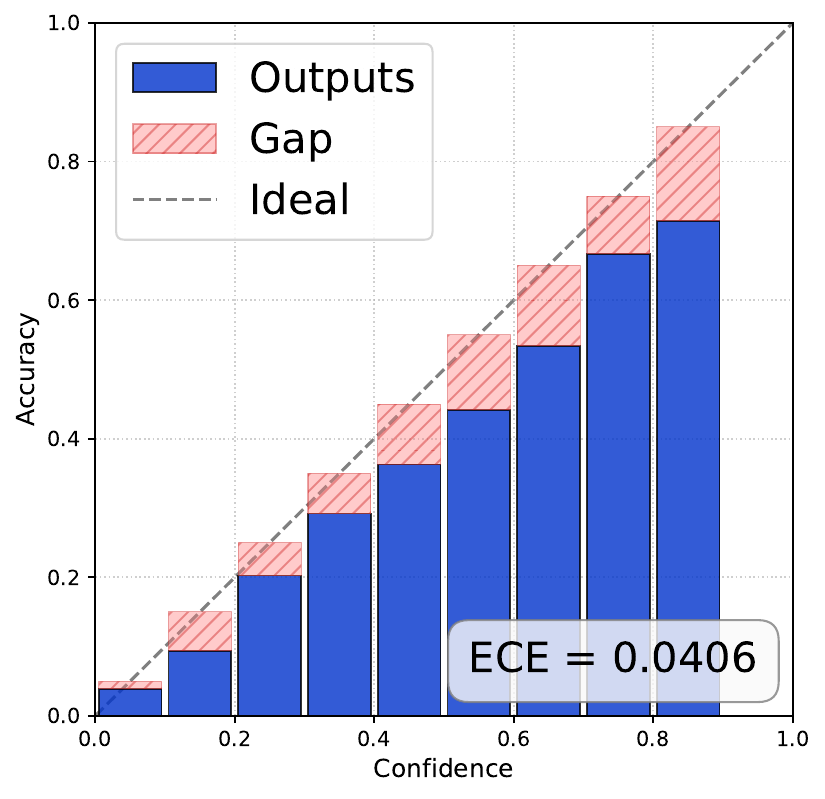}
    }
    \subfigure[FedGMT-v2]{%
        \includegraphics[width=0.185\textwidth]{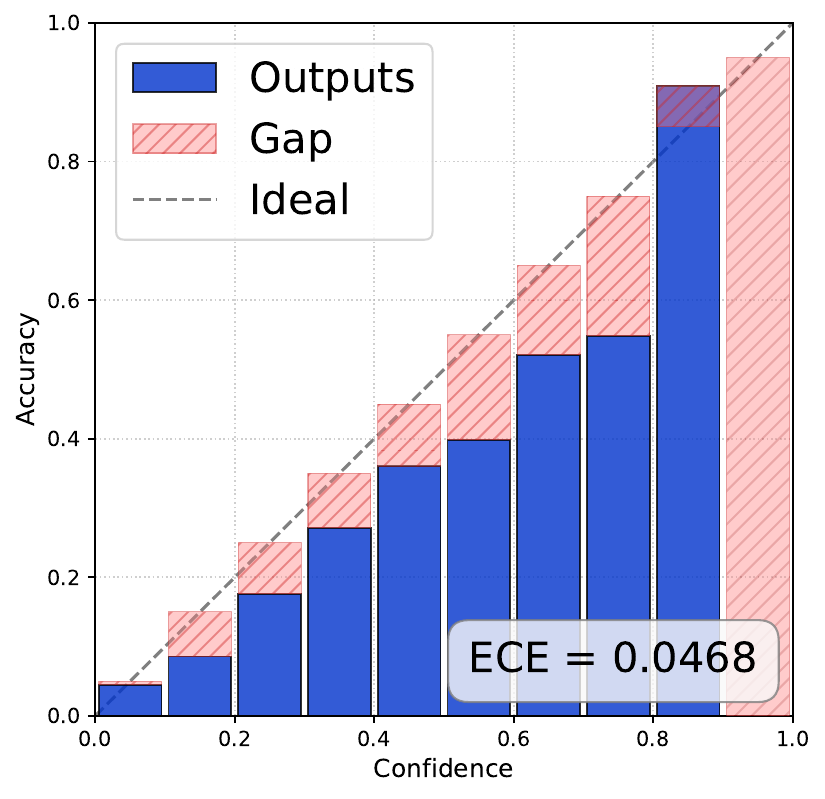}
    }
    \subfigure[Proposed]{%
        \includegraphics[width=0.185\textwidth]{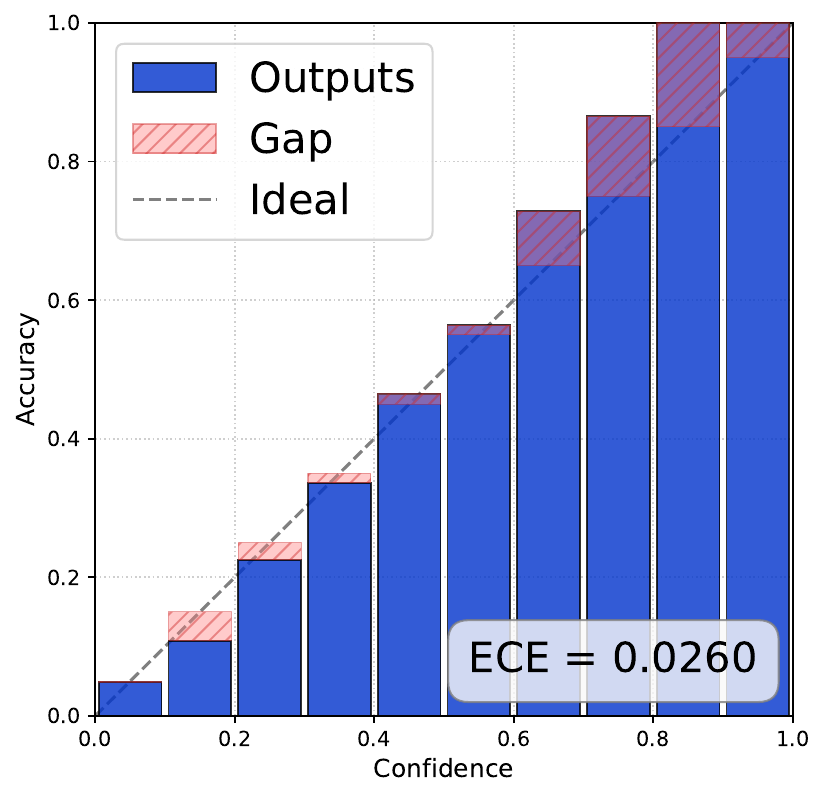}
    }

    {\centering \small (b) CIFAR-100\par}
    \caption{Comparison of reliability diagrams for different FL methods}\label{fig:RD}
\end{figure*}

Figure~\ref{fig:VaryingSizes} illustrates the training progress of the proposed framework and benchmark methods for varying local dataset sizes. Overall, increasing the local dataset size tends to improve convergence and test accuracy, although the extent of the improvement varies across methods and datasets. In the MNIST single-class setting, the gains are relatively limited because additional samples do not increase local class diversity. In contrast, more pronounced improvements are observed on CIFAR-100, where larger local datasets better represent each device's underlying class distribution and yield lower-variance, more reliable local updates. Although larger local datasets improve the peak performance of the SAM-based methods, they do not eliminate their post-peak degradation, particularly for FedGMT-v2 on CIFAR-100.
Consistent with the observations in Fig.~\ref{fig:VaryingLabels}, the proposed framework exhibits the most stable convergence behavior across the considered dataset sizes. This stability can be attributed to its uncertainty-aware distribution-level aggregation, which reduces the influence of less reliable local posterior estimates, particularly when local data are scarce.

Figure~\ref{fig:PowerBudget} illustrates the training progress of the proposed framework and benchmark methods under varying transmission power budgets at the edge devices. 
As the power budget decreases, all methods exhibit slower convergence and lower test accuracy because the devices must more strongly attenuate their local updates to satisfy the power constraints, thereby increasing aggregation distortion. This degradation is more pronounced on CIFAR-100, where learning is more sensitive to reduced update fidelity. Compared with the benchmark methods, the proposed framework exhibits relatively small performance degradation across the considered power budgets. This behavior is consistent with the combined effects of distribution-level aggregation and the proposed power-control strategy, which minimizes an upper bound on the aggregation distortion. Moreover, the proposed framework maintains its attained accuracy at all considered power levels, demonstrating stable convergence under power-constrained aggregation.

\subsection{Calibration Performance}

In scenarios where inference performance is constrained by scarce and heterogeneous data, calibration performance plays a critical role in ensuring the reliability of model predictions. 
In other words, the predicted probabilities produced by the model should be trustworthy and should accurately reflect the true likelihood of each outcome. 
For example, in the classification problem, the model’s confidence score, which is typically represented by the largest softmax output, should correspond closely to the actual accuracy observed for samples assigned that level of confidence.
In this subsection, we evaluate the calibration performance  by presenting both reliability diagrams and the expected calibration error (ECE) \cite{guo2017calibration}. 

The reliability diagram illustrates the relationship between predicted confidence and empirical accuracy, while the ECE quantifies their average discrepancy. To construct the reliability diagram and compute the ECE, we partition $N_\text{test}=10{,}000$ test samples into 10 bins based on predicted confidence scores. For each bin $\mathcal{B}_j$, the average confidence and accuracy are defined as
\begin{align}
\text{conf}(\mathcal{B}_j) &= \frac{1}{|\mathcal{B}_j|} \sum_{i \in \mathcal{B}_j} \hat{p}_i, \label{eq:conf}
\end{align}
\begin{align}
\text{acc}(\mathcal{B}_j) &= \frac{1}{|\mathcal{B}_j|} \sum_{i \in \mathcal{B}_j} \mathbb{1}(l_i = \hat{l}_i), \label{eq:acc}
\end{align}
where $\hat{p}_i$ denotes the predicted confidence score for sample $i$, $\hat{l}_i$ is the predicted label, and $l_i$ is the ground-truth label.

In the reliability diagram, the empirical accuracy $\text{acc}(\mathcal{B}_j)$ for each bin $j \in \{0, 1, \ldots, 9\}$ is plotted against the corresponding average predicted confidence.
The ECE is defined as
\begin{align}
    \text{ECE} = \sum_{j=0}^{9} \frac{|\mathcal{B}_j|}{N_\text{test}} \left| \text{acc}(\mathcal{B}_j) - \text{conf}(\mathcal{B}_j) \right|.
\end{align}

Although several non-Bayesian calibration techniques, such as post-hoc temperature scaling \cite{guo2017calibration}, have been proposed to improve confidence calibration, their objective is solely to adjust prediction confidence after training, which differs from the main objective of the proposed framework, namely enabling stable learning under data scarcity and heterogeneous data distributions. 
Furthermore, post-hoc calibration methods require a sufficiently large validation set to reliably estimate calibration parameters, which is difficult to secure in the considered wireless FL scenario where each device possesses only a limited local dataset. 
Therefore, these methods are not included in our comparison.

Figure~\ref{fig:RD} presents the reliability diagrams and ECEs of the models trained with the proposed and benchmark FL methods under the default simulation setup, as detailed in Tables~\ref{table:SimulSetup} and~\ref{table:HyperParam}.
It is observed that the benchmark FL methods, which are based on a frequentist learning approach, exhibit large gaps between the bars and the diagonal.
In particular, the frequentist approach tends to produce overconfident predictions, as indicated by these gaps.
In contrast, the proposed framework achieves noticeably smaller gaps than the benchmark methods, owing to its distribution-level aggregation and probabilistic inference.
Furthermore, the proposed method achieves an ECE less than half that of most benchmark methods.

\section{Conclusion}
In this paper, we proposed a distribution-level AirComp framework for Bayesian FL in wireless edge networks with data scarcity and heterogeneity. By maintaining posterior distributions, the proposed approach captures model uncertainty and mitigates local overfitting and client drift, while enabling efficient aggregation of local posteriors over wireless channels. We further developed a transmit power control strategy based on convergence analysis under wireless impairments, showing that communication-induced distortion can become a major convergence-limiting factor under constrained power budgets. Simulation results demonstrate improved test accuracy, convergence stability, and calibration performance over conventional FL methods. Overall, the proposed framework provides a robust and communication-efficient solution for wireless FL under limited and heterogeneous local data.

\appendices
\section{Proof of Theorem \ref{thr:CostReduction}} \label{app:prrof_thm1}
For simplicity, we consider $E=1$ and $d=F$, which implies $N_1 = N_2 = 1$. Consequently, the superscripts $(n_1)$ and $(n_2)$ that denote the OFDM symbol indices are omitted in the subsequent convergence analysis. 
Under Assumption 1, the decrease in the cost function during the first phase of training round $t$ is bounded above as 
\begin{align}\label{eq:phase1_update}
    &L\!\left(\boldsymbol{\nu}_t, \boldsymbol{\rho}_{t+1}\right) - L\!\left(\boldsymbol{\nu}_t, \boldsymbol{\rho}_{t}\right) \nonumber\\
    &\le \nabla_\rho L\!\left(\boldsymbol{\nu}_t, \boldsymbol{\rho}_{t}\right)^\intercal \left( \boldsymbol{\rho}_{t+1} - \boldsymbol{\rho}_{t}\right) + \frac{\Lambda_\rho}{2} \left\Vert \boldsymbol{\rho}_{t+1} - \boldsymbol{\rho}_{t} \right\Vert^2 \nonumber\\
    &\overset{(a)}{=} \nabla_\rho L\!\left(\boldsymbol{\mu}_t, \boldsymbol{\rho}_{t}\right)^\intercal \left( -\eta \sum_{k\in\mathcal{K}}\pi_k\nabla_\rho L_k\left(\boldsymbol{\mu}_{t},\boldsymbol{\rho}_{t}\right)+\eta\boldsymbol{\xi}_{\rho_t} + \mathbf{z}'_{\rho_t} \right)\nonumber\\
    &~~~ + \frac{\Lambda_\rho}{2} \left\Vert -\eta \sum_{k\in\mathcal{K}}\pi_k\nabla_\rho L_k\left(\boldsymbol{\mu}_{t},\boldsymbol{\rho}_{t}\right)+\eta\boldsymbol{\xi}_{\rho_t} + \mathbf{z}'_{\rho_t} \right\Vert^2\nonumber\\
    &=-\eta \left\Vert \nabla_\rho L\!\left(\boldsymbol{\mu}_t, \boldsymbol{\rho}_{t}\right)\right\Vert^2 + \nabla_\rho L\!\left(\boldsymbol{\mu}_t, \boldsymbol{\rho}_{t}\right)^\intercal \left( \eta\boldsymbol{\xi}_{\rho_t} + \mathbf{z}'_{\rho_t} \right)\nonumber\\
    &~~~ + \frac{\Lambda_\rho}{2} \left\Vert -\eta \nabla_\rho L\!\left(\boldsymbol{\mu}_t, \boldsymbol{\rho}_{t}\right)+\eta\boldsymbol{\xi}_{\rho_t} + \mathbf{z}'_{\rho_t} \right\Vert^2,
\end{align}
where the equality $(a)$ comes from the fact that $\mathbf{\Delta}_{\rho_{t,k}}= -\eta \nabla_{\rho}\frac{L_k(\boldsymbol{\nu}_t, \boldsymbol{\rho}_t)}{\bar{\delta}}$ for $E=1$, and $\mathbf{z}'_{\rho_t}\sim\mathcal{CN}\Big(0, \frac{\bar{\delta}_{\rho_t}\sigma_z^2}{\gamma}\mathbf{I}_d\Big)$.
Using \eqref{eq:phase1_update}, the expected decrease in the cost function over channel noise is upper bounded as
\begin{align}
    &\mathbb{E}_{\mathbf{z}}\!\left[ L\!\left(\boldsymbol{\nu}_t, \boldsymbol{\rho}_{t+1}\right) - L\!\left(\boldsymbol{\nu}_t, \boldsymbol{\rho}_{t}\right) \right] \nonumber\\
    &\le -\eta \left\Vert \nabla_\rho L\!\left(\boldsymbol{\mu}_t, \boldsymbol{\rho}_{t}\right)\right\Vert^2 + \eta\nabla_\rho L\!\left(\boldsymbol{\mu}_t, \boldsymbol{\rho}_{t}\right)^\intercal \boldsymbol{\xi}_{\rho_t} \nonumber\\
    &~~~+ \frac{\Lambda_\rho}{2} \left\Vert  -\eta \nabla_\rho L\!\left(\boldsymbol{\mu}_t, \boldsymbol{\rho}_{t}\right)+\eta\boldsymbol{\xi}_{\rho_t}\right\Vert^2 + \frac{\Lambda_\rho d \bar{\delta}_{\rho_t}\sigma_z^2}{2\gamma} \nonumber\\
    &\le -\eta \left\Vert \nabla_\rho L\!\left(\boldsymbol{\mu}_t, \boldsymbol{\rho}_{t}\right)\right\Vert^2 + \eta\nabla_\rho L\!\left(\boldsymbol{\mu}_t, \boldsymbol{\rho}_{t}\right)^\intercal \boldsymbol{\xi}_{\rho_t}\nonumber\\
    &~~~+ \frac{\Lambda_\rho\eta^2}{2} \left(\left\Vert  \nabla_\rho L\!\left(\boldsymbol{\mu}_t, \boldsymbol{\rho}_{t}\right) \right\Vert^2 + \left\Vert\boldsymbol{\xi}_{\rho_t}\right\Vert^2\right) +\frac{\Lambda_\rho d \bar{\delta}_{\rho_t}\sigma_z^2}{2\gamma} \nonumber
    \end{align}
\begin{align}
    &\overset{(a)}{\le} -\eta \left\Vert \nabla_\rho L\!\left(\boldsymbol{\mu}_t, \boldsymbol{\rho}_{t}\right)\right\Vert^2 + \frac{\eta}{2}\left(\left\Vert\nabla_\rho L\!\left(\boldsymbol{\mu}_t, \boldsymbol{\rho}_{t}\right)\right\Vert^2 + \left\Vert\boldsymbol{\xi}_{\rho_t}\right\Vert^2 \right) \nonumber\\
    &~~~+ \frac{\Lambda_\rho\eta^2}{2} \left(\left\Vert \nabla_\rho L\!\left(\boldsymbol{\mu}_t, \boldsymbol{\rho}_{t}\right) \right\Vert^2 + \left\Vert\boldsymbol{\xi}_{\rho_t}\right\Vert^2\right)  +\frac{\Lambda_\rho d \bar{\delta}_{\rho_t}\sigma_z^2}{2\gamma}\nonumber\\
    &= -\frac{\eta}{2}\left(1-\Lambda_\rho\eta\right) \left\Vert \nabla_\rho L\!\left(\boldsymbol{\mu}_t, \boldsymbol{\rho}_{t}\right)\right\Vert^2 +\frac{\eta}{2}\left(1+\Lambda_\rho\eta\right)   \left\Vert\boldsymbol{\xi}_{\rho_t}\right\Vert^2\nonumber\\
    &~~~+\frac{\Lambda_\rho d \bar{\delta}_{\rho_t}\sigma_z^2}{2\gamma}, \label{eq:bound1}
\end{align}
where the inequality $(a)$ comes from the inequality of arithmetic and geometric means.

Using $\mathbf{\Sigma}_{t+1}=\text{diag}\!\left(\boldsymbol{\rho}_{t+1}\right)^{-1}$, an upper bound on the decrease in the cost function during the second phase can be derived  similarly to \eqref{eq:phase1_update}, as
\begin{align}\label{eq:update_phase2}
    &L\!\left(\boldsymbol{\nu}_{t+1}, \boldsymbol{\rho}_{t+1}\right) - L\!\left(\boldsymbol{\nu}_{t}, \boldsymbol{\rho}_{t+1}\right)\nonumber\\
    &\le \nabla_\nu L\!\left(\boldsymbol{\nu}_{t}, \text{diag}\big(\mathbf{\Sigma}_{t+1}^{-1}\big)\right)^\intercal \left( \boldsymbol{\nu}_{t+1} - \boldsymbol{\nu}_{t}\right) + \frac{\Lambda_\nu}{2} \left\Vert \boldsymbol{\nu}_{t+1} - \boldsymbol{\nu}_{t} \right\Vert^2 \nonumber\\
    &=-\eta \left\Vert \nabla_\nu L\!\left(\boldsymbol{\nu}_{t}, \text{diag}\big(\mathbf{\Sigma}_{t+1}^{-1}\big)\right)\right\Vert^2 + \nabla_\nu L\!\left(\boldsymbol{\nu}_{t}, \text{diag}\big(\mathbf{\Sigma}_{t+1}^{-1}\big)\right)^\intercal\nonumber\\
    &~~~\times\!\left( \eta\boldsymbol{\xi}_{\nu_t} \!+\! \mathbf{z}'_{\nu,t} \right)\! +\! \frac{\Lambda_\nu}{2}\! \left\Vert -\eta \nabla_\nu L\!\left(\boldsymbol{\nu}_{t}, \text{diag}\big(\mathbf{\Sigma}_{t+1}^{-1}\big)\right)\!+\!\eta\boldsymbol{\xi}_{\nu_t} \!+\! \mathbf{z}'_{\nu_t} \right\Vert^2.
\end{align}
From \eqref{eq:update_phase2}, the expected decrease in the cost function over channel noise can be derived in a manner similar to \eqref{eq:bound1}, as 
\begin{align}
    &\mathbb{E}_{\mathbf{z}}\!\left[ L\!\left(\boldsymbol{\nu}_{t}, \text{diag}\big(\mathbf{\Sigma}_{t+1}^{-1}\big)\right) - L\!\left(\boldsymbol{\nu}_{t}, \text{diag}\big(\mathbf{\Sigma}_{t+1}^{-1}\big)\right)  \right] \nonumber\\
    &\le -\frac{\eta}{2}\left(1-\Lambda_\nu\eta\right) \left\Vert \nabla_\nu L\!\left(\boldsymbol{\nu}_{t}, \text{diag}\big(\mathbf{\Sigma}_{t+1}^{-1}\big)\right)\right\Vert^2 \nonumber\\
    &~~~+ \frac{\eta}{2}\left(1 +\Lambda_\nu\eta\right) \left\Vert\boldsymbol{\xi}_{\nu_t}\right\Vert^2+ \frac{\Lambda_\nu d \bar{\delta}_{\nu_t}\sigma_z^2}{2\gamma}. \label{eq:bound2}
\end{align}

Set $\eta=\frac{1}{\sqrt{T}}$, and extend the expectation over randomness in the trajectory across $T$ training rounds as follows:

\begin{align}
    &\mathbb{E}\!\left[L\!\left(\boldsymbol{\nu}_0, \boldsymbol{\rho}_{0}\right) \!-\! L\!\left(\boldsymbol{\nu}_T, \boldsymbol{\rho}_T\right)\right] = \mathbb{E}\bigg[\! \sum_{t=0}^{T-1} L\!\left(\boldsymbol{\nu}_t, \boldsymbol{\rho}_t\right) \!-\! L\!\left(\boldsymbol{\nu}_{t+1}, \boldsymbol{\rho}_{t+1}\right)\! \bigg] \nonumber\\
    &= \mathbb{E}\bigg[ \sum_{t=0}^{T-1} L\!\left(\boldsymbol{\nu}_t, \boldsymbol{\rho}_t\right) - L\!\left(\boldsymbol{\nu}_t, \boldsymbol{\rho}_{t+1}\right) \nonumber\\
    &\qquad\qquad+ L\!\left(\boldsymbol{\nu}_t, \boldsymbol{\rho}_{t+1}\right) - L\!\left(\boldsymbol{\nu}_{t+1}, \boldsymbol{\rho}_{t+1}\right) \bigg] \nonumber\\
    &\overset{(a)}{\ge}  \mathbb{E}\bigg[\sum_{t=0}^{T-1} \frac{1}{2\sqrt{T}}\left(1-\frac{\Lambda_\rho}{\sqrt{T}}\right) \left\Vert \nabla_\rho L\!\left(\boldsymbol{\mu}_t, \boldsymbol{\rho}_t\right)\right\Vert^2 \nonumber\\
    &\qquad\qquad-\frac{1}{2\sqrt{T}}\left(1+\frac{\Lambda_\rho}{\sqrt{T}}\right)   \left\Vert\boldsymbol{\xi}_{\rho_t}\right\Vert^2 - \frac{\Lambda_\rho d \bar{\delta}_{\rho_t}\sigma_z^2}{2\gamma}  \nonumber\\
    &\qquad\qquad+\frac{1}{2\sqrt{T}}\left(1-\frac{\Lambda_\nu}{\sqrt{T}}\right) \left\Vert \nabla_\nu L\!\left(\boldsymbol{\nu}_{t}, \text{diag}\big(\mathbf{\Sigma}_{t+1}^{-1}\big)\right)\right\Vert^2\nonumber\\
    &\qquad\qquad- \frac{1}{2\sqrt{T}}\left(1 + \frac{\Lambda_\nu}{\sqrt{T}}\right) \left\Vert\boldsymbol{\xi}_{\nu_t}\right\Vert^2 - \frac{\Lambda_\nu d \bar{\delta}_{\nu_t}\sigma_z^2}{2\gamma}\bigg] \nonumber\\
    &\overset{(b)}{=}  \frac{1}{2\sqrt{T}}\mathbb{E}\bigg[ \sum_{t=0}^{T-1}\left(1-\frac{\Lambda_\rho}{\sqrt{T}}\right) \left\Vert \nabla_\rho L\!\left(\boldsymbol{\mu}_t, \boldsymbol{\rho}_t\right)\right\Vert^2\nonumber\\
    &\qquad\qquad\qquad+ \left(1-\frac{\Lambda_\nu}{\sqrt{T}}\right) \left\Vert \nabla_\nu L\!\left(\boldsymbol{\nu}_{t}, \text{diag}\big(\mathbf{\Sigma}_{t+1}^{-1}\big)\right)\right\Vert^2 \nonumber\\
    &\qquad\qquad\qquad-\left(1+\frac{\Lambda_\rho}{\sqrt{T}}\right)   \left\Vert\boldsymbol{\xi}_{\rho_t}\right\Vert^2 - \left(1 + \frac{\Lambda_\nu}{\sqrt{T}}\right) \left\Vert\boldsymbol{\xi}_{\nu_t}\right\Vert^2 \bigg]\nonumber
\end{align}
\begin{align}
    &~~~- \frac{\Lambda_\rho \sigma_z^2}{2\gamma} \mathbb{E}\left[ \frac{1}{T}\sum_{t=0}^{T-1}\sum_{k\in\mathcal{K}}\pi_k\left\Vert \nabla_\rho  L_k\!\left(\boldsymbol{\mu}_{t}, \boldsymbol{\rho}_{t}\right) \right\Vert^2 \right]\nonumber\\
    &~~~- \frac{\Lambda_\nu \sigma_z^2}{2\gamma } \mathbb{E}\left[ \frac{1}{T}\sum_{t=0}^{T-1}\sum_{k\in\mathcal{K}}\pi_k\!\left\Vert \nabla_\nu  L_k\!\left(\boldsymbol{\nu}_{t}, \text{diag}\big(\mathbf{\Sigma}_{t+1}^{-1}\big)\right) \right\Vert^2 \right] \nonumber\\
    &=\frac{1}{2\sqrt{T}}\mathbb{E}\bigg[ \sum_{t=0}^{T-1}\left(1-\frac{\Lambda_\rho}{\sqrt{T}}\right) \left\Vert \nabla_\rho L\!\left(\boldsymbol{\mu}_t, \boldsymbol{\rho}_t\right)\right\Vert^2\nonumber\\
    &\qquad\qquad\qquad+\left(1-\frac{\Lambda_\nu}{\sqrt{T}}\right) \left\Vert \nabla_\nu L\!\left(\boldsymbol{\nu}_{t}, \text{diag}\big(\mathbf{\Sigma}_{t+1}^{-1}\big)\right)\right\Vert^2\bigg] \nonumber\\
    &~~~ - \frac{1}{2\sqrt{T}}\mathbb{E}\bigg[\sum_{t=0}^{T-1}\left(1\!+\!\frac{\Lambda_\rho}{\sqrt{T}}\right)   \left\Vert\boldsymbol{\xi}_{\rho_t}\right\Vert^2 + \left(1 \!+\! \frac{\Lambda_\nu}{\sqrt{T}}\right) \left\Vert\boldsymbol{\xi}_{\nu_t}\right\Vert^2 \bigg]\nonumber\\
    &~~~ - \frac{\Lambda_\rho \sigma_z^2}{2\gamma T} \mathbb{E}\bigg[\sum_{t=0}^{T-1} \sum_{k\in\mathcal{K}}\pi_k\left\Vert \nabla_\rho  L_k\!\left(\boldsymbol{\mu}_{t}, \boldsymbol{\rho}_{t}\right) \right\Vert^2 \bigg] \nonumber\\
    &~~~- \frac{\Lambda_\nu \sigma_z^2}{2\gamma T} \mathbb{E}\bigg[ \sum_{t=0}^{T-1}\!\sum_{k\in\mathcal{K}}\!\pi_k\!\left\Vert \nabla_\nu  L_k\!\left(\boldsymbol{\nu}_{t}, \text{diag}\big(\mathbf{\Sigma}_{t+1}^{-1}\big)\right) \right\Vert^2 \!\bigg], \label{eq:telescope}
\end{align}
where the inequality $(a)$ comes from \eqref{eq:bound1} and \eqref{eq:bound2}, and the equality $(b)$ comes from  $\bar{\delta}_{\rho_t}=\sum_{k\in\mathcal{K}} \frac{\pi_k}{d} \Vert \eta\nabla_\rho  L_k(\boldsymbol{\mu}_{t}, \boldsymbol{\rho}_{t}) \Vert^2$ and $\bar{\delta}_{\nu_t}=\sum_{k\in\mathcal{K}} \frac{\pi_k}{d} \big\Vert \eta\nabla_\nu  L_k\big(\boldsymbol{\nu}_{t}, \text{diag}\big(\mathbf{\Sigma}_{t+1}^{-1}\big)\big) \big\Vert^2$ for $E=1$.

\ifCLASSOPTIONcaptionsoff
  \newpage
\fi

\bibliographystyle{IEEEtran}
\bibliography{bibliography}

\end{document}